\documentclass[journal]{IEEEtran}

\usepackage{adjustbox}
\usepackage{makecell}
\usepackage{multirow}
\usepackage{enumitem}

\newcommand{\sit}[1]{\textrm{\textit{#1}}}
\usepackage{tcolorbox}
\tcbuselibrary{skins, breakable}
\usepackage{amsmath}
\usepackage{amssymb}

\usepackage[hyphens]{url} % line breaks in URLs
\usepackage[breaklinks=true]{hyperref}

\newtcolorbox{rqanswerbox}{
  enhanced,
  colback=gray!15,
  colframe=gray!15,                       % 让主边框“伪装成背景色”
  boxrule=0pt,                            % 主边框宽度为0
  borderline west={2pt}{0pt}{gray!70},   % 左侧竖线（宽度和颜色可调）
  sharp corners,
  boxsep=4pt,
  left=4pt, right=4pt, top=4pt, bottom=4pt,
  before skip=1ex,
  after skip=1ex
}

\ifCLASSINFOpdf
\else
\fi
\usepackage{array}
\begin{document}
%
% paper title
% Titles are generally capitalized except for words such as a, an, and, as,
% at, but, by, for, in, nor, of, on, or, the, to and up, which are usually
% not capitalized unless they are the first or last word of the title.
% Linebreaks \\ can be used within to get better formatting as desired.
% Do not put math or special symbols in the title.
\title{Improving LLM-Based Go Code Review through Issue-List Generation and Context Augmentation}
%
%
% author names and IEEE memberships
% note positions of commas and nonbreaking spaces ( ~ ) LaTeX will not break
% a structure at a ~ so this keeps an author's name from being broken across
% two lines.
% use \thanks{} to gain access to the first footnote area
% a separate \thanks must be used for each paragraph as LaTeX2e's \thanks
% was not built to handle multiple paragraphs
%

\author{Kexin Sun, Yucong Guan, Jiaqi Sun, Hongyu Kuang, Guoping Rong, Dong Shao, He Zhang, Xiaoxing Ma, and Christoph Treude% <-this % stops a space}
\thanks{K. Sun, Y. Guan, J. Sun, H. Kuang, G. Rong, D. Shao, H. Zhang, and X. Ma, are with State Key Lab for Novel Software Technology, Nanjing University, Nanjing, China (e-mail: kexinsun@smail.nju.edu.cn; yucongguan@smail.nju.edu.cn; jiaqisun@smail.nju.edu.cn; khy@nju.edu.cn; ronggp@nju.edu.cn; dongshao@nju.edu.cn; hezhang@nju.edu.cn; xxm@nju.edu.cn).}%
\thanks{C. Treude is with School of Computing and Information Systems, Singapore Management University, Singapore (e-mail: ctreude@smu.edu.sg).}
\thanks{Corresponding author: Hongyu Kuang (e-mail: khy@nju.edu.cn).}}

% note the % following the last \IEEEmembership and also \thanks -
% these prevent an unwanted space from occurring between the last author name
% and the end of the author line. i.e., if you had this:
%
% \author{....lastname \thanks{...} \thanks{...} }
%                     ^------------^------------^----Do not want these spaces!
%
% a space would be appended to the last name and could cause every name on that
% line to be shifted left slightly. This is one of those "LaTeX things". For
% instance, "\textbf{A} \textbf{B}" will typeset as "A B" not "AB". To get
% "AB" then you have to do: "\textbf{A}\textbf{B}"
% \thanks is no different in this regard, so shield the last } of each \thanks
% that ends a line with a % and do not let a space in before the next \thanks.
% Spaces after \IEEEmembership other than the last one are OK (and needed) as
% you are supposed to have spaces between the names. For what it is worth,
% this is a minor point as most people would not even notice if the said evil
% space somehow managed to creep in.

% The paper headers
\markboth{Journal of \LaTeX\ Class Files,~Vol.~14, No.~8, August~2015}%
{Shell \MakeLowercase{\textit{et al.}}: Bare Demo of IEEEtran.cls for IEEE Journals}
% The only time the second header will appear is for the odd numbered pages
% after the title page when using the twoside option.
%
% *** Note that you probably will NOT want to include the author's ***
% *** name in the headers of peer review papers.                   ***
% You can use \ifCLASSOPTIONpeerreview for conditional compilation here if
% you desire.

% If you want to put a publisher's ID mark on the page you can do it like
% this:
%\IEEEpubid{0000--0000/00\$00.00~\copyright~2015 IEEE}
% Remember, if you use this you must call \IEEEpubidadjcol in the second
% column for its text to clear the IEEEpubid mark.

% use for special paper notices
%\IEEEspecialpapernotice{(Invited Paper)}

% make the title area
\maketitle

% As a general rule, do not put math, special symbols or citations
% in the abstract or keywords.
\begin{abstract}
Large language models (LLMs) have shown strong potential for automating code review, yet their practical utility depends heavily on the design of generation and context strategies.
In this paper, we investigate how to improve LLM-based code review through generation strategy and contextual augmentation.
We first propose an \textit{issue-list review} paradigm, in which LLMs enumerate all potential issues rather than reporting only the single most important one (i.e., \textit{primary-issue review}).
We then systematically compare three types of code context augmentation --- neighboring, LSP-based semantics, and IR-based similar co-change context --- and study how they influence issue discovery.
Finally, we integrate candidates from no-context and context-enhanced generation to improve review coverage, and introduce refinement-guided pruning to keep the candidate list at a practical size.
We evaluate our approach on 1,438 Go review instances and use downstream code refinement as the main effectiveness metric, i.e., how often the generated candidate list contains at least one comment that induces the same code change as the final human revision.
% The results show that issue-list review improves over primary-issue review from 17.15\% to 21.83\% (+4.68\%).
% Among the contextual augmentation strategies, combining neighboring context with IR-based similar context yields the best context-enhanced setting, reaching 25.59\% (+3.76\%).
% Merging its candidate list with no-context candidates further improves the result to 28.00\% (+2.41\%), substantially outperforming CodeReviewer (15.02\%) and moving closer to the human-oracle ceiling of 36.09\%.
For comparison, we also evaluate comments generated by CodeReviewer, a model trained
specifically for review comment generation, as well as ground-truth human review comments (which serve as a practical upper bound), under the same refinement-based evaluation.
The results show that our best configuration, combining issue-list review, neighboring and similar co-change context, and candidate integration, reaches 28.00\% refinement exact match, a statistically significant overall gain of +10.85 percentage points over primary-issue review without any additional context (17.15\%), substantially outperforming CodeReviewer (15.02\%) and moving closer to the human-oracle ceiling of 36.09\%.
Our refinement-guided pruning strategy then reduces the average candidate count from 7.2 to 3.1 at top-5 while retaining nearly the full benefit, making the candidate list easier to inspect.
% show total improvement to be strong
\end{abstract}

% Note that keywords are not normally used for peerreview papers.

% For peer review papers, you can put extra information on the cover
% page as needed:
% \ifCLASSOPTIONpeerreview
% \begin{center} \bfseries EDICS Category: 3-BBND \end{center}
% \fi
%
% For peerreview papers, this IEEEtran command inserts a page break and
% creates the second title. It will be ignored for other modes.
\IEEEpeerreviewmaketitle

%%%%%%%%%%%%%%%%%%%%%%%%%%%%%%%%%%%%%%%% Introduction %%%%%%%%%%%%%%%%%%%%%%%%%%%%%%%%%%%%%%%%
\section{Introduction}
Code review is a fundamental activity in modern software engineering~\cite{rigby2013convergent}.
It helps expose functional defects early in the development cycle while identifying valuable refactoring opportunities and enhancing system evolvability~\cite{mantyla2008types,DBLP:conf/msr/BellerBZJ14,turzo2024makes}.
Accordingly, code review plays a critical role in safeguarding software quality~\cite{mcintosh2016empirical}, mitigating the accumulation of long-term technical debt~\cite{fu2022potential}, and facilitating knowledge transferring within development teams~\cite{bacchelli2013expectations}.
However, as software projects continue to grow in scale and complexity, manual code review has become increasingly time-consuming and knowledge-intensive, placing a substantial burden on developers~\cite{zhou2017scalability}.

To assist and accelerate this process, prior work has explored a wide range of automation techniques, ranging from early static analysis tools that detect potentially problematic code patterns~\cite{ayewah2008using}, to recent learning-based approaches~\cite{li2022automating,tufano2022using} for generating review comments and for generating revised code based on the review feedback (i.e., code refinement).
The rise of large language models (LLMs) has brought unprecedented opportunities for further improvement~\cite{hou2024large}.
Many open-source communities and tech companies have started deploying LLM-based reviewers in practical development environments.
However, these practical deployments also demonstrate that the actual utility of AI-assisted reviews is highly dependent on their design.
Low-quality review comments may even increase the burden on developers~\cite{cihan2025automated, sun2026does}.
% This necessitates the exploration of more effective generation and contextual strategies to ensure that LLM-based code reviews are concise, specific, and practically actionable.
%generate comprehensive issue reports based on an established defect taxonomy
Therefore, in this paper, we investigate how to improve LLM-based code review by prompting LLMs to comprehensively report potential issues and by supplying the models with the carefully curated code contexts.

%TO DO: I found these two references from the emails to support our observation, but I'm not sure if they are enough. Do we need to do some sampling to further justify it?
In practice, human reviewers often write each review comment around one primary concern, rather than listing every potential issue~\cite{chong2021assessing,googleReviewerLookingFor}.
By manually inspecting 100 randomly sampled human-review comments, we confirmed this situation by finding that 97\% of comments center on a single issue.
Thus, we started replicating this \sit{primary-issue review} style by prompting LLMs to report the only issue that is considered to be the most important for each reviewed hunk.
However, we then found that this prioritization constraint may cause the LLM to miss issues that it could otherwise identify due to its ``primary concern'' not aligned with human reviewers.
To better understand how LLM discover review issues, we choose to alternatively prompt LLMs to ``explore'' as many potential issues as possible, then examine which of these align with actual developer concerns.
We refer to this as \sit{issue-list review}.

\begin{figure}[t]
	\centering
    \setlength{\abovecaptionskip}{3pt}
	\includegraphics[width=0.38\textwidth]{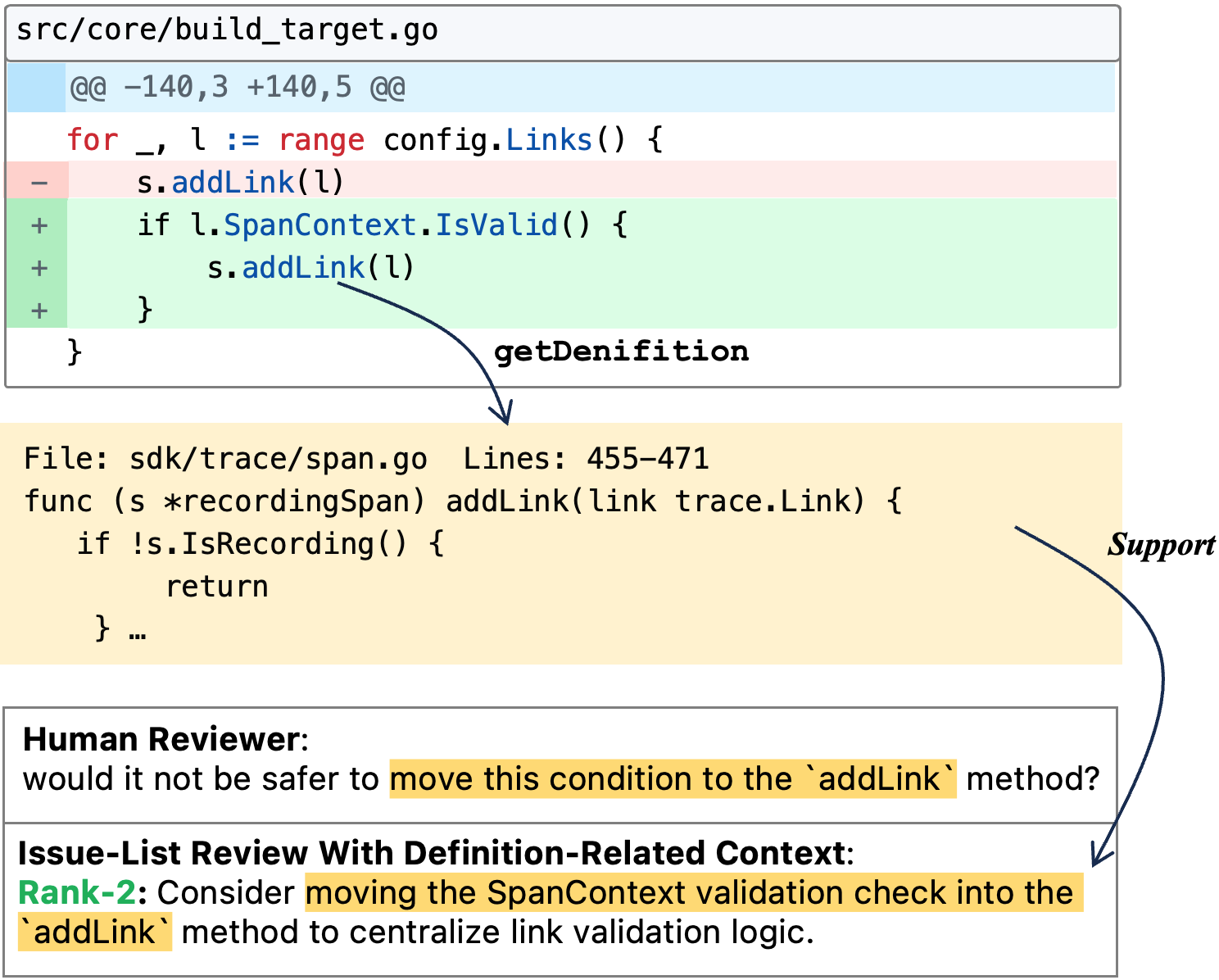}
	\caption{An example of how definition context assists LLM in code review.}
	\label{fig:denifition_example}
\end{figure}

\begin{figure}[t]
	\centering
    \setlength{\abovecaptionskip}{3pt}
	\includegraphics[width=0.44\textwidth]{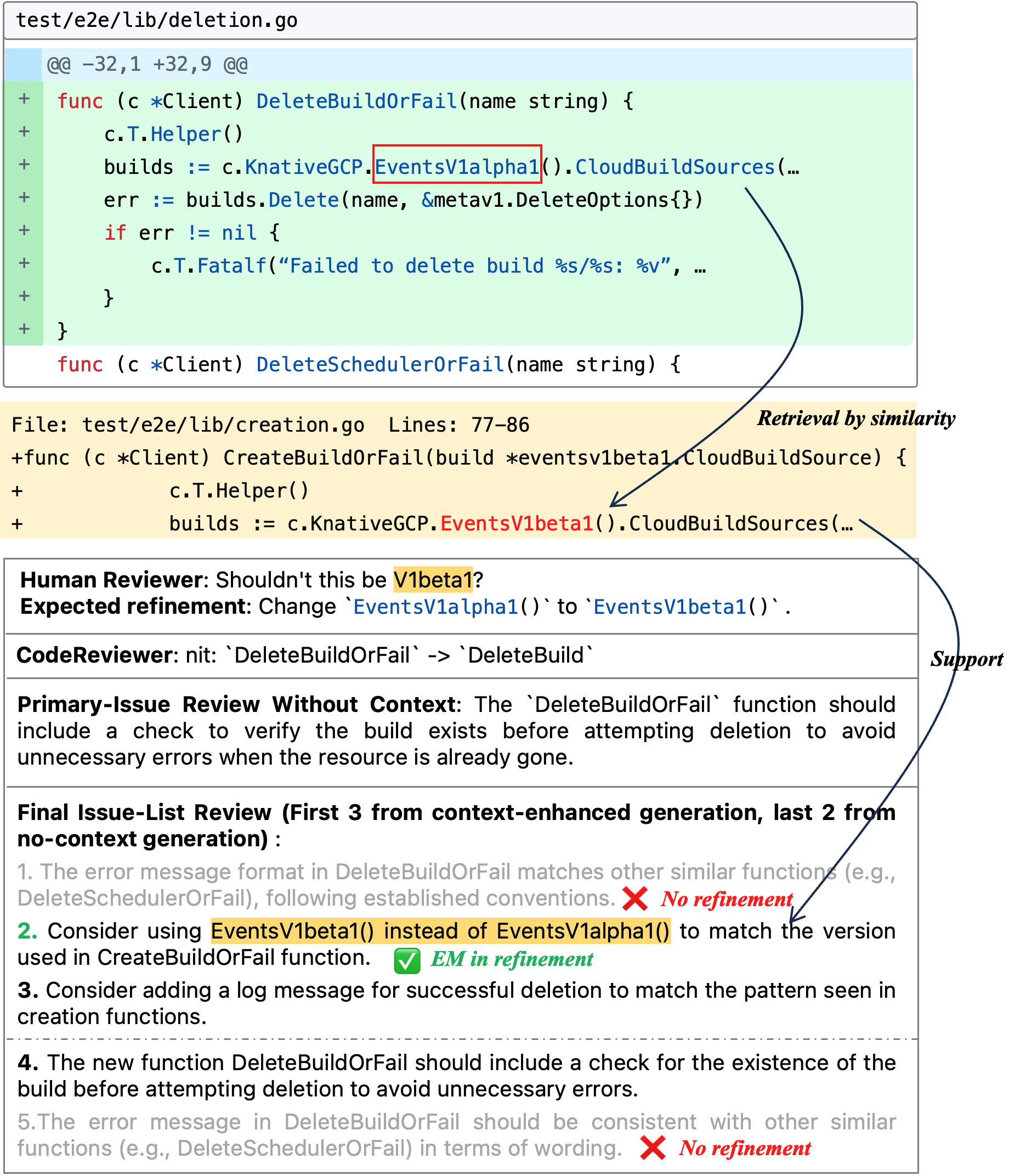}
	\caption{A motivating example on how issue-list generation, context augmentation, and refinement-guided pruning improve generated review comments.}
	\label{fig:motivation_example}
\end{figure}

At the same time, the local diff alone is often insufficient for code review.
Useful review comments may depend on broader code evidence.
For example, in Figure~\ref{fig:denifition_example}, the definition of \texttt{addLink} could  help LLMs recognize that a validity check is best placed inside the function itself rather than at the call site.
Similarly, comparable implementations retrieved from co-modified files could allow models to spot inconsistencies between the current hunk and other code involved in the same development task (shown in Figure~\ref{fig:motivation_example}, we further detail these cases in subsequent sections.)
Both cases highlight that the richer code context, which lies beyond the reviewed hunk, matters for effective LLM-based code review.

Motivated by this, we investigate how different contextual augmentation strategies influence LLM-based code review.
Specifically, we consider three types of code context:
(1) \textit{neighboring context}, i.e. the enclosing function body of the reviewed hunk, which helps the model understand the local control and data flow of the change;
(2) \textit{semantic context} obtained through Language Server Protocol (LSP)-based program navigation, which enables the model to inspect structurally relevant elements via specific LSP queries, such as resolving the underlying implementation of an identifier via \texttt{getDefinition} or tracking its usage across the codebase via \texttt{getReferences};
(3) \textit{similar co-change context}, i.e., similar patterns retrieved via IR (Information Retrieval)-based search from co-modified files, which helps the model compare the current change against related implementations involved in the same development task.
The first strategy extends context through spatial locality around the reviewed hunk.
The remaining two go beyond such locality by retrieving context through structural relevance (via program navigation) and code similarity (via IR-based search).
We also checked whether these three contexts are complementary, and found that more context is not always beneficial.
%We also checked whether combining these context yields further improvements, and found that more context is not always beneficial.
% Since neighboring context provides the most immediate information for understanding the reviewed change, we treat it as the foundation to examine whether the remaining two strategies—which go beyond the spatial locality around the reviewed hunk via program navigation and IR-based search—can further improve review quality, individually or combined.
%Our results suggest that more context is not always beneficial.
In particular, supplementing the combination of neighboring and similar contexts with semantic context failed to bring further performance gains.
%and even led to a slight decline.
This observation indicates that, rather than simply expanding the model’s focus, additional context may also blind the model to issues it would otherwise catch.

% This complementarity between basic and context-enhanced generation suggests that no single strategy is sufficient on its own.
% By merging candidates from both settings, we can recover useful reviews that either would miss in isolation.
% However, naive aggregation quickly inflates the candidate set.
% We therefore introduce a refinement-guided filtering step.
% We use CodeReviewer~\cite{li2022automating}, a well-performing pre-trained model for code refinement, to simulate the likely code changes in response to candidate comments, and then remove candidates unlikely to trigger meaningful refinement and deduplicate those leading to similar refinements.
% This allows us to retain candidates that induce diverse code changes, while keeping the candidate set manageable.

To avoid missing issues that no-context generation (i.e., generation without any additional code context) may explore, we further merge candidates from the no-context setting and the best context-enhanced setting into a unified pool.
However, the naive aggregation inevitably inflates the candidate set.
We thus introduce a refinement-guided filtering step.
Specifically, we introduced CodeReviewer~\cite{li2022automating}, a well-performing pre-trained model for further code refinement, to simulate the likely code changes in response to candidate comments.
We then filtered candidates that fail to trigger any code revision and de-duplicate those leading to similar refinements, allowing us to retain candidates that induce diverse revisions, while keeping the candidate set manageable.

% We evaluate our approach on 1,438 Go review instances reconstructed from the CodeReviewer benchmark~\cite{li2022automating}.
% We use performance on the downstream code refinement task as our primary metric, since it directly reflects the practical impact of the review comments, i.e.\ the proportion of cases in which the generated candidate list contains at least one comment that induces the same code change as the final human revision.
We evaluate our approach on 1,438 Go review instances reconstructed from the CodeReviewer benchmark~\cite{li2022automating}.
We focus on Go because it generally enables more reliable static symbol resolution than dynamically typed languages such as Python, and avoids the heavy build configuration required by languages such as Java, making it a practical choice to fully exploit the benefits of LSP-based context retrieval.
Due to the general inefficiency of text-similarity metrics (e.g., BLEU) to assess the actual quality of review comments~\cite{lu2025deepcrceval, naik2025crscore}, we set our primary metric as the performance on downstream code refinement tasks.
Specifically, we measure how often the generated comment(s) contain at least one comment that induces an \textbf{E}xact-\textbf{M}atch code change with the final human revision.
We refer to this metric as \sit{RefineEM}.
For comparison, we also evaluate comments generated by CodeReviewer, which is trained to generally produce one comment per hunk during code-review generation.
We also evaluate ground-truth comments from human review under the same refinement-based evaluation as a practical upper bound.
Our main observations are as follows.
Without any additional context, issue-list review is able to reach 21.83\% \sit{RefineEM}, significantly improving over primary-issue review (17.15\% \sit{RefineEM}) by +4.68 percentage points (pp), while increasing the average number of candidates from 1.0 to 3.1 per case.
This observation suggests that the bottleneck lies not only in issue identification itself, but also in whether LLMs assign sufficient priority to explore the identified issues in the final review output.
Among contextual augmentation strategies, combining neighboring context with IR-based similar co-change context is the best-performing strategy, reaching 25.59\% \sit{RefineEM} and improving over the no-context setting (21.83\%) by +3.76 pp.
% Adding LSP-based semantic context does not further help and can even hurt performance, suggesting that context shifts the model's review focus rather than uniformly expanding it.
Merging candidates from this best context-enhanced generation with no-context candidates further improves the result to \textbf{28.00\%} \sit{RefineEM} (+2.41 pp), showing \textbf{an overall gain of +10.85 pp} over primary-issue review without any context.
That is, our approach outperforms CodeReviewer by 15.02\% \sit{RefineEM}, and is closer to the human-oracle ceiling of 36.09\% \sit{RefineEM}.
Furthermore, to achieve this gain with less candidate comments (i.e., 7.2 comments per case on average), we proposed the refinement-guided pruning strategy to reduce the average candidates to 3.1 at top-5 while retaining the achieved refinement, making the candidate list easier to inspect.

This paper makes the following contributions:
\begin{itemize}[itemsep=1pt, topsep=3pt, leftmargin=15pt]
  \item We propose an \textbf{issue-list review} paradigm for LLM-based code review and show that it \textbf{improves} downstream refinement effectiveness \textbf{over primary-issue review}.
  \item We systematically compare \textbf{three types of code context augmentation} for LLM-based code review, covering neighboring, LSP-based semantic, and IR-based similar co-change context, and analyze how different forms of context influence issue discovery.
  \item We merge candidates from no-context and context-enhanced generation to maximize review coverage, and introduce \textbf{refinement-guided pruning} to control candidate explosion while preserving nearly full refinement benefits.
  \item We evaluate the proposed strategy on 1,438 Go review instances using downstream code refinement as the primary metric, and show that it outperforms existing baselines.
\end{itemize}

Our code and data are publicly available~\cite{replicationPackage} (to be published on Zenodo after acceptance).

%%%%%%%%%%%%%%%%%%%%%%%%%%%%%%%%%%%%%%%% Motivation Example %%%%%%%%%%%%%%%%%%%%%%%%%%%%%%%%%%%%%%%%
\section{Motivation Example}

% During code review, a human reviewer typically provides a natural-language comment on the reviewed hunk to identify concerns or suggest improvements.
% The developer may then address the comment by revising the code, which we refer to as \sit{refinement}.
% In this section, we use a \href{https://github.com/google/knative-gcp/pull/1381#discussion_r451066167}{case} adapted from \texttt{google/knative-gcp} (Figure~\ref{fig:motivation_example}) to illustrate how issue-list generation and code context can help LLMs better automate this process by generating comprehensive and specific review comments.

In this section, we detail a \href{https://github.com/google/knative-gcp/pull/1381#discussion_r451066167}{case} adapted from \texttt{google/knative-gcp} (Figure~\ref{fig:motivation_example}) to illustrate how issue-list generation and code context augmentation help LLMs generate comprehensive and specific review comments.
The reviewed hunk introduces a test helper function \texttt{DeleteBuildOrFail()}, which calls the \texttt{EventsV1alpha1()} API.
The human reviewer identified a single concern: \textit{``Shouldn't this be V1beta1?''}, suggesting that the newer \texttt{EventsV1beta1()} API should be used instead, while our baseline approach, CodeReviewer, only produces a superficial naming suggestion (\textit{``nit: \texttt{DeleteBuildOrFail} $\rightarrow$ \texttt{DeleteBuild}''}), missing the API version issue entirely.

When no additional context is provided and the LLM is asked to report only the single most important issue, it focuses on defensive programming: \textit{``The \texttt{DeleteBuildOrFail} function should include a check to verify the build exists before attempting deletion.''}.
Without broader context, the model can only reason about the local hunk and defaults to generic robustness suggestions.
Even when we relax the prompt to issue-list generation under the same no-context setting (comments~\#4 and~\#5 in Figure~\ref{fig:motivation_example}), the model still prioritizes this defensive-programming issue, while only additionally mentioning that the error message should be consistent with similar functions.

However, when we augment the prompt with the similar co-change patterns, in this case \texttt{test/e2e/lib/creation.go}, which uses \texttt{EventsV1beta1()}, the model recognizes that the reviewed hunk uses the outdated \texttt{EventsV1alpha1()} API and raises the human-aligned concern---\textit{``using \texttt{EventsV1beta1()} instead of \texttt{EventsV1alpha1()}''}---as its second-ranked suggestion (comment~\#2).
This demonstrates that both issue-list generation and code context augmentation help LLMs surface the issues that human reviewers actually prioritize.

To further improve coverage, we aggregate comments from both context-enhanced and no-context generation into a unified pool
(the context-enhanced generation may occasionally miss issues that no-context generation would surface; we discuss such a case in Figure~\ref{fig:combination_fail_example_2}).
However, naive merging inflates the candidate set: the unified pool in Figure~\ref{fig:motivation_example} contains five candidates.
We therefore use the refinements induced by candidate comments to prune vague candidates and deduplicate candidates that lead to similar revisions.
For example, comments~\#1 and~\#5 both raise the same error-message consistency concern, yet neither specifies a concrete change and neither triggers any refinement.
Our refinement-guided pruning strategy removes such redundant and low-impact candidates (shown greyed out in Figure~\ref{fig:motivation_example}), reducing the list to three candidates and elevating the human-aligned suggestion to the top position.

As this example illustrates, issue-list generation and context augmentation broaden the coverage of potential defects, while refinement-guided pruning helps remove less actionable candidates from the final list.
Together, they form an effective approach for LLM-based code review.

%%%%%%%%%%%%%%%%%%%%%%%%%%%%%%%%%%%%%%%% Approach %%%%%%%%%%%%%%%%%%%%%%%%%%%%%%%%%%%%%%%%

\begin{figure*}[t]
	\centering
    \setlength{\abovecaptionskip}{3pt}
	\includegraphics[width=0.98\textwidth]{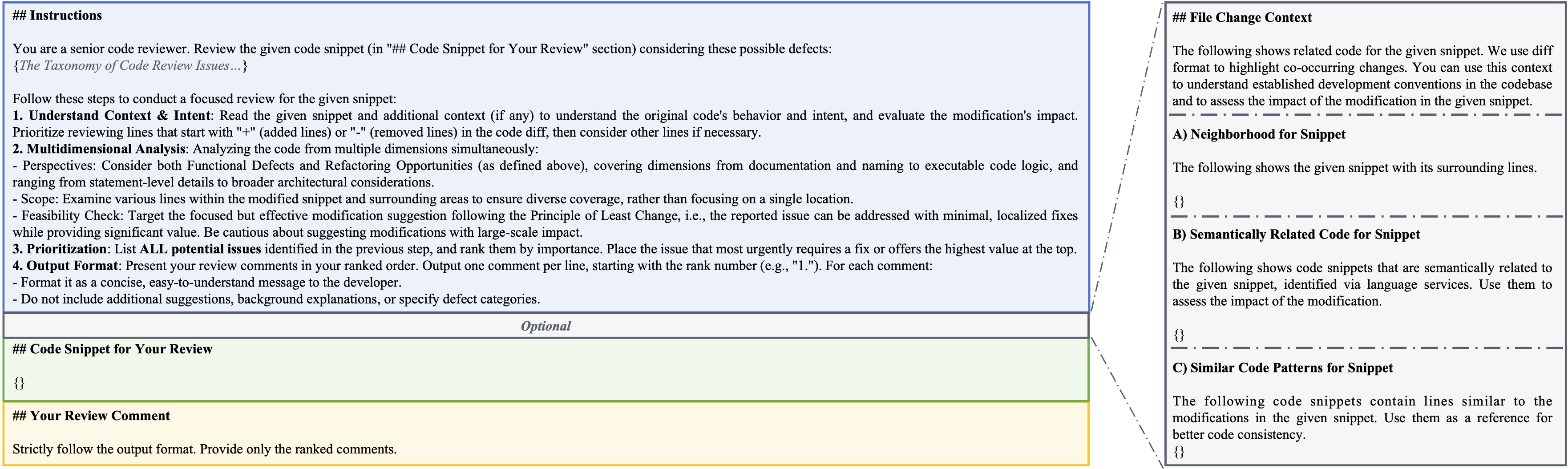}
	\caption{Prompt template used for issue-list code review, with optional slots for neighboring, semantic, and similar-code context.}
	\label{fig:prompt}
\end{figure*}

\begin{table*}[t]
\caption{\textbf{Taxonomy of Code Review Issues} Used in the Review Prompt}
\label{tab:taxonomy}
\centering
\scriptsize
\begin{tabular}{>{\centering\arraybackslash}p{0.06\textwidth} >{\centering\arraybackslash}p{0.15\textwidth} p{0.70\textwidth}}
\hline
\textbf{Category} & \textbf{Defect Type} & \multicolumn{1}{c}{\textbf{Description}} \\
\hline

\multirow{7}{*}[-0.5em]{Functional} & Functional Defect & Missing or incorrectly implemented functionality requiring code additions or larger modifications to the existing solution. \\

 & Logic Defect & Defects made with comparison operations, control flow, and computations and other types of logical mistakes. \\

 & Check Defect & Validation mistakes or mistakes made when detecting an invalid value. \\

 & Resource Defect & Defects related to variable initialization, system memory management, and manipulating or releasing data or other resources.\\

 & Interface Defect & Mistakes made when interacting with other parts of the software, such as an existing code library, a hardware device, a database, or an operating system. \\

 & Timing Defect & Concurrency issues in multi-threaded applications involving shared resources. \\

 & Support Defect & Incorrect configurations or version issues for support systems or libraries. \\
\hline

\multirow{6}{*}[-1.5em]{Refactoring} & Solution Approach Defect & A wide range of defects that truly represent the Alternative Approach in its most fruitful form, including the defects: Semantic Duplication, Semantic Dead Code, Change Function, Use Standard Method, New Functionality (subtype definitions omitted for brevity; see the prompt for details).\\

 & Organization Defect & Structural issues including Move Functionality, Long Sub-routine, Dead Code, Duplication, Complex Code, Statement Issues, and Consistency. \\

 & Alternate Output & Comments that suggest modifying the error message, toast message, alert, or change what is returned by a function. \\

 & Naming Convention & Violations of identifier naming conventions. \\

 & Visual Representation & Whitespace, blank lines, code rearrangements, and indentation-related comments. \\

 & Documentation & Comments to add/modify comments or documentation to aid code comprehension. \\
\hline
\end{tabular}
\end{table*}

\section{Approach}

%TO DO:  Figure~\ref{fig:overview}
% Figure~\ref{fig:overview} provides an overview of our LLM-based code review approach.
Our LLM-based code review approach proceeds in three stages.
Given a diff hunk, we first prompt the LLM to list all potential issues (\S\ref{sec:reviewer}).
We then investigate three types of code contextual augmentation:
(1) neighboring context around the review hunk,
(2) semantically related code context obtained through LSP-based program navigation,
and (3) similar co-change context through IR-based search (\S\ref{sec:context}).
Finally, we combine candidates across different context settings and apply refinement-based pruning to the candidate set (\S\ref{sec:filter}).
% Finally, to maximize effectiveness, we combine candidates across different context settings and apply refinement-based pruning to the candidate set, discarding comments that fail to trigger any code refinement and merging those that lead to the same changes (\S\ref{sec:filter}).

\subsection{Issue-List Review}
\label{sec:reviewer}
To help the LLM discover review issues more comprehensively, our review-guidance prompt consists of two components: a taxonomy of code review issues and a structured review workflow.

The taxonomy, summarized in Table~\ref{tab:taxonomy}, is drawn from prior empirical studies~\cite{mantyla2008types, DBLP:conf/msr/BellerBZJ14,turzo2024makes} that investigate the types of issues commonly identified during manual code review.
In general, these issues can be categorized into two high-level types:
\textit{Functional} issues, which may lead to incorrect behavior or system failure when the code is executed, and \textit{Refactoring} issues, which make the code less compliant with standards, more error-prone, or more difficult to modify, extend, or understand.
We use this taxonomy as a checklist to guide the model toward a broader and more systematic search over potential review concerns.

The review workflow then instructs the LLM to first understand the intent of the change and any provided context, then systematically analyze the diff against each issue category, rank all identified issues by urgency or value, and finally present them as concise, easy-to-understand comments in ranked order.
The complete prompt is shown in Figure~\ref{fig:prompt}.
% The complete prompt is available in our replication package.

\subsection{Contextual Augmentation}
\label{sec:context}
To better comprehend the target hunk, we utilize the language server to parse the post-change repository snapshot.
This allows us to obtain structured program analysis results through LSP queries.
In particular, our approach mainly relies on three types of queries:
\begin{itemize}[itemsep=1pt, topsep=3pt, leftmargin=15pt]
    \item \texttt{documentSymbol}, which returns the list of symbols and their ranges in a file;
    \item \texttt{getDefinition}, which resolves a symbol to its declaration site;
    \item \texttt{getReferences}, which locates all usage sites of a symbol across the codebase.
\end{itemize}
The following three context augmentation strategies are all based on these language-server capabilities.

\vspace{1ex}
\subsubsection{Function-Level Neighboring Context}
We treat the function as the natural unit of code comprehension.
To retrieve functions related to the reviewed hunk, we use the \texttt{documentSymbol} query to obtain the symbol list of the reviewed file.
We then check whether any modified line in the hunk falls within the range of a function symbol;
if so, the complete body of the matched function is included in the neighboring context.
If no enclosing function is found, or if the language server fails to parse the file, we fall back to a line-based window of up to 50 lines centered on the hunk.

\vspace{1ex}
\subsubsection{LSP-Based Semantic Context}
To gather semantically related context, we design a set of query rules to determine which LSP queries are executed for the reviewed hunk.
Each query returns a set of code locations, which we further expand into context blocks according to the query type.
The resulting blocks are then deduplicated and ranked to form the final semantic context.

\vspace{1ex}
\textit{2.1) LSP Query Rules.}
We design an identifier-exhaustive query strategy over the modified lines.
Specifically, we use tree-sitter~\cite{tree-sitter} to extract all identifiers appearing in the changed lines of the post-change file, and execute both \texttt{getDefinition} (to locate what the symbol denotes) and \texttt{getReferences} (to trace how it is used across the codebase) for each identifier.
This strategy provides a broad view of how the modified code is connected to the rest of the codebase.

When executing the query plan, we automatically augment each \texttt{getDefinition} query with a companion \texttt{getTypeDefinition} query on the same symbol.
While \texttt{getDefinition} locates the declaration of a symbol, \texttt{getTypeDefinition} further reveals the definition of its underlying type.
For example, if a modification introduces a new use of a variable \texttt{cfg}, \texttt{getDefinition} may lead us to its declaration site such as ``\texttt{var cfg Config}'', while \texttt{getTypeDefinition} further reveals the definition of
\texttt{Config} itself, which provides a more complete definition context.

\vspace{1ex}
\textit{2.2) Query-Based Block Expansion.}
Each LSP query returns a list of locations, each pinpointing a single line in a file.
We expand each location into a context block according to the query type.
For \texttt{getDefinition} and \texttt{getTypeDefinition} results, we aim to retrieve the complete definition of the target symbol.
For type- or function-level symbols, we further perform \texttt{documentSymbol} query on the destination file to resolve the symbol's full bounding range.
For variable-level symbols, whose definitions typically span a single line, we instead expand with a fixed window of up to three non-blank lines above and below the anchor.                                                                       For \texttt{getReferences} and \texttt{getSearch} results, which similarly resolve to single-line locations, we apply the same window-based expansion.
After this step, each query is associated with a list of context blocks, each
represented as \sit{block}(\sit{file}, \sit{anchor}, \sit{span}, \sit{context}).

\vspace{1ex}
\textit{2.3) Block Organization and Deduplication.}
We first deduplicate query commands by their resolved locations:
if two queries return identical results, we keep only the one with the earlier anchor position (i.e., smaller (\sit{query.line}, \sit{query.cursor})).
This is because queries are planned from individual changed symbols and rules without pre-merging; we defer merging until confirming that they resolve to the same locations.

We then merge the blocks produced by individual queries.
For \texttt{getReferences} and \texttt{getSearch} queries, a single symbol may yield many nearby locations.
We therefore merge any overlapping blocks from the same query into a single larger block.
Additionally, we apply the same merging to blocks produced by \texttt{getDefinition} and \texttt{getTypeDefinition}: since a reviewed change may reference both a type and its members, the resulting definition blocks often overlap.
We therefore collect all such definition-related blocks, rank them by span size, and discard any block whose \sit{block.anchor} falls within an already-retained larger block.

Furthermore, we organize blocks across different target symbols and queries.
We group queries by target symbol; within each group, queries are ordered by type:
\texttt{getDefinition} $>$ \texttt{getTypeDefinition} $>$ \texttt{getReferences} $>$ \texttt{getSearch}.
The groups are then sequenced by the first appearance of each symbol in the hunk.
We then iterate over all blocks following this order, checking each \sit{block} against those already admitted into the context (including any pre-existing neighboring-context block).
For blocks produced by \texttt{getDefinition} and \texttt{getTypeDefinition}, if \sit{block.anchor}
falls within an existing block, the block is discarded (because even after definition-level deduplication, a block may still overlap with the neighboring-context block);
otherwise, if the block only partially overlaps with existing context, we trim the overlapping portion and keep the side containing \sit{block.anchor}.
For \texttt{getReferences} and \texttt{getSearch} blocks, we rank candidates by their file's edit distance to the reviewed file, then examine them in that order and discard any block whose \sit{block.anchor} is already covered by existing context.
We retain at most $k{=}5$ blocks for each such query to prevent high-frequency symbols from overwhelming the context.

These steps reduce redundant information for the limited context budget.
Finally, each retained \sit{block} is rendered using the following template:
\begin{tcolorbox}[
  width=0.48\textwidth,
  center,
  colback=white,       % 纯白底色（相当于无底色）
  colframe=black,      % 纯黑边框
  boxrule=0.4pt,       % 极细边框（0.4pt 是 LaTeX 标准线宽）
  sharp corners,       % 直角（去除圆边）
  left=3pt, right=3pt, top=3pt, bottom=3pt, % 缩小内部文字到边框的间距
  before skip=4pt, after skip=4pt           % 缩小盒子与上下正文的间距
]
\footnotesize\ttfamily
[query header]\\
File: [\sit{block.file}]\quad Lines: [\sit{block.span}]\\
{[\sit{block.content}]}
\end{tcolorbox}
where the query header is derived from \sit{query.type} and \sit{query.symbol}, which explains how the block relates to the reviewed hunk:
\begin{itemize}[itemsep=1pt, topsep=3pt, leftmargin=15pt]
  \item\texttt{getDefinition}: ``Find the definition of `\sit{symbol}''';
  \item\texttt{getTypeDefinition}: ``Find the type definition of `\sit{symbol}' (which is `\sit{type}')'';
  \item\texttt{getReferences}: ``Find the reference to `\sit{symbol}''';
  \item\texttt{getSearch}: ``Find the existing occurrence of `\sit{symbol}'''.
\end{itemize}

\vspace{1ex}
\subsubsection{IR-Based Similar Co-Change Context}
To construct similar co-change context, we take files co-modified in the same commit as the retrieval scope, as these files are naturally coupled by the same development task and thus provide a focused search space for related implementation patterns.
We then partition these files into candidate blocks, score each by its similarity to the modified lines, and select the top-$N$ as the final similar context.

\vspace{1ex}
\textit{3.1) Syntax-Guided Code Partitioning.}
To avoid arbitrary line-based slicing, we partition each candidate file according to its syntactic structure.
By querying \texttt{documentSymbol} for symbol ranges within the co-modified files, we
divide each file into \sit{function blocks} (areas covered by function-level symbols) and \sit{inter-function blocks} (fragments between recognized functions).
To prevent excessive fragmentation, any inter-function blocks whose line count does not exceed $\tau_{\text{merge}}$ are merged into the nearest adjacent function block.
Conversely, oversized blocks are further partitioned to maintain an ideal context granularity of $S_{\text{target}}$ lines.
Specifically, if a block's line count $s$ exceeds $S_{\text{target}}$, we aim to split it into $N = \lfloor s / S_{\text{target}} \rfloor$ sub-blocks.
To preserve semantic integrity, we prioritize placing cut points at blank lines, which serve as natural boundaries in source code.
Furthermore, to avoid extreme splitting, we constrain each sub-block's size fall within $[S_{\min}, S_{\max}]$ lines.
We formalize this partitioning process as a dynamic programming problem that minimizes the following cost:
\[
\footnotesize
  \sum_{k=1}^{N} \bigl[(s_k - S_{\text{target}})^2 + \lambda_{\text{cut}} \cdot \delta(k)\bigr]
  \quad \text{subject to } S_{\min} \le s_k \le S_{\max},
  \label{eq:dp-partition}
\]
where $s_k$ is the size of the $k$-th sub-block, and $\delta(k) = 1$ is an indicator that equals $1$ if the $k$-th cut is forced onto a non-blank line, and $0$ otherwise.
The penalty parameter $\lambda_{\text{cut}}$ implies that as long as the nearest blank line deviates from $S_{\text{target}}$ by no more than $\sqrt{\lambda_{\text{cut}}}$ lines, the algorithm will tolerate this size deviation to preserve the natural boundary.
Together, these steps yield a set of candidate blocks with consistent granularity and well-defined semantic scope.
By default, we set $S_{\text{target}}\!=\!20$, $S_{\min}\!=\!10$, $S_{\max}\!=\!30$, $\tau_{\text{merge}}\!=\!10$, and $\lambda_{\text{cut}}\!=\!50$.

\vspace{1ex}
\textit{3.2) Change-Based Similarity Scoring.}
To identify candidate blocks that share implementation patterns with the reviewed change, we score each candidate block against the changed lines in the hunk.
We use the added lines (prefixed with \texttt{+}) as target lines; if the hunk contains only deletions, we fall back to the deleted lines.
Both target lines and candidate block lines are tokenized using a subword tokenizer that splits on underscores and CamelCase boundaries and lowercases all tokens.
We then represent these lines as TF-IDF vectors over unigram-to-trigram features, fitting the vocabulary jointly over all target and candidate lines.
Given a candidate \textit{block}, its similarity score is computed as follows:
\[
  \mathrm{Score}(\textit{block}) = \sum_{i=1}^{m} \tilde{w}_i \cdot \max_{l \in \textit{block}} \mathrm{Sim}(t_i, l),
\]
where $\mathrm{Sim}(t_i, l)$ denotes the cosine similarity between target line $t_i$ and candidate line $l$.
The $\max$ over $l \in \textit{block}$ captures the localized relevance, i.e. a candidate block is valuable as long as it contains at least one highly similar line that provides a concrete reference for that implementation pattern.
The normalized weight $\tilde{w}_i = w_i / \sum_j w_j$, where $w_i = \|\mathbf{t}_i\|_2$ is the $\ell_2$ norm of $t_i$'s TF-IDF vector, rewarding similarity to informationally rich target lines over generic ones.
If $t_i$ is a comment line, its base weight is further discounted by a factor of $0.3$ (prior to normalization) to prioritize executable code over natural language description.

\vspace{1ex}
\textit{3.3) Candidate Selection and Ranking.}
% After scoring, we first apply a similarity filter $\mathcal{F}_{\text{sim}}$ (default: $\mathrm{Score}(\textit{block}) > 0$) to exclude irrelevant candidates, and further discard blocks whose line span overlaps with the neighboring context, since the two are partitioned differently and would otherwise introduce redundant content.
After scoring, we first apply a similarity filter $\mathcal{F}_{\text{sim}}$ (default: $\mathrm{Score}(\textit{block}) > 0$) to exclude irrelevant candidates, and further discard blocks whose line span overlaps with any existing context block to avoid redundancy.
We then rank the remaining candidates primarily by their similarity score, with ties broken by the file-path edit distance between \sit{block.file} and the reviewed file.
Finally, we select the top $5$ blocks as the final similar context, each rendered in ranked order as ``\texttt{File: [\sit{block.file}] Lines: [\sit{block.span}]\textbackslash n[\sit{block.content}]}''.

\vspace{1ex}
Once the context blocks have been selected, we apply a unified diff-aware rendering step to all three context sources.
For each selected block, we check whether its line span intersects with any changed region from the same review batch.
If so, the corresponding lines are replaced in-place with the unified-diff fragment, preserving the ``\texttt{@@}'' header as well as the ``\texttt{+}''/``\texttt{-}'' prefixes.
This allows the model to see both the surrounding unchanged code and the exact modification within a single coherent view.

\subsection{Refinement-Guided Candidate Integration}
\label{sec:filter}
Different generation settings may surface different but complementary review comments.
We therefore integrate candidates from a context-enhanced setting and a no-context setting to maximize coverage, while using a refinement-guided step to control candidate-set growth.
We use $\sit{Cands}_{\sit{ctx}}=\langle c^{\sit{ctx}}_1,\dots,c^{\sit{ctx}}_m\rangle$ to denote the ranked comments generated with contextual augmentation, and $\sit{Cands}_{\sit{base}}=\langle c^{\sit{base}}_1,\dots,c^{\sit{base}}_n\rangle$ to denote those generated without additional context.
We construct a unified candidate pool by concatenating them: $\sit{Cands}_{\sit{pool}}\!=\!\sit{Cands}_{\sit{ctx}}\!\oplus\!\sit{Cands}_{\sit{base}}$, placing the context-enhanced candidates first to reflect their higher overall effectiveness.

To prune $\sit{Cands}_{\sit{pool}}$, we use CodeReviewer~\cite{li2022automating}, a well-performing pre-trained model for code refinement, to simulate the revised code each candidate comment would induce.
This design follows an effect-based intuition: rather than judging a candidate by its surface text, we assess it by the concrete code edit it is predicted to trigger.
For each candidate $c_i \in \sit{Cands}_{\sit{pool}}$, we use $\sit{ref}_i$ to denote its predicted refinement.
We then iterate through $\sit{Cands}_{\sit{pool}}$ in order and apply two pruning rules:
(1) if $\sit{ref}_i$ is identical to the pre-change code, $c_i$ is discarded as it fails to induce any meaningful revision;
(2) if $\sit{ref}_i$ matches the predicted refinement $\sit{ref}_j$ of a higher-ranked candidate $c_j$ ($j\!<\!i$), $c_i$ is discarded as redundant.
The remaining candidates form the final review comment set.

%%%%%%%%%%%%%%%%%%%%%%%%%%%%%%%%%%%%%%%% Experimental Setup %%%%%%%%%%%%%%%%%%%%%%%%%%%%%%%%%%%%%%%%
\section{Experimental Setup}
In this section, we introduce the research questions, dataset, and evaluation metrics used in our study.

\subsection{Research Questions}
We structure our evaluation around the following research questions, which examine the effectiveness of issue-list review, contextual augmentation, and candidate integration, respectively.
These questions follow a progressive structure:
RQ1 establishes the issue-list review baseline,
RQ2 investigates how different context augmentation strategies improve upon it,
and RQ3 examines how candidates from the best context-enhanced and no-context settings can be further combined.
We also conduct an ablation study to quantify the contribution of each design choice in the full pipeline in \S\ref{sec:ablation}.  

\begin{itemize}[itemsep=1pt, topsep=3pt, leftmargin=15pt]
    \item \textbf{RQ1}: How does issue-list review affect the effectiveness of LLM-based code review?
    \item \textbf{RQ2}: How do different contextual augmentation strategies affect the effectiveness of LLM-generated code reviews?
    To make this analysis easier to follow, we break RQ2 into four sub-questions:
    \begin{itemize}[itemsep=1pt, topsep=2pt, leftmargin=15pt]
        \item \textbf{RQ2a}: How effective is neighboring context for issue-list review?
        \item \textbf{RQ2b}: How effective is LSP-based semantic context for issue-list review?
        \item \textbf{RQ2c}: How effective is IR-based similar co-change context for issue-list review?
        \item \textbf{RQ2d}: How can multiple context types be combined for optimal performance?
    \end{itemize}
    \item \textbf{RQ3}: How can candidates from different context settings be combined to maximize effectiveness while maintaining practical candidate sizes?
\end{itemize}

To answer \textbf{RQ1}, we compare issue-list review (denoted as \textbf{\sit{List-None}}) against primary-issue review (denoted as \textbf{\sit{Primary-None}}) under the same no-context setting, where the LLM is given only the reviewed hunk without any additional code context.
In \sit{Primary-None}, we keep the prompt in Figure~\ref{fig:prompt} unchanged except for the review workflow: instead of prompting the model to list all identified issues and rank them, we prompt it to ``\textit{point out the \textbf{single most significant issue}---the one that most urgently requires a fix or offers the highest value if addressed.}''

As additional comparisons, we include \textbf{\sit{CodeReviewer}}, trained on the original review comment generation dataset, as a supervised baseline for review comment generation.
We also report its downstream refinement performance by feeding its generated comments into the same refinement model used in our evaluation.
Finally, we report the refinement performance induced by ground-truth human review comments (denoted as \textbf{\sit{Human-Oracle}}) as a practical upper-bound reference.

To answer \textbf{RQ2}, we investigate how different contextual augmentation strategies affect review effectiveness, using \textbf{\sit{List-None}} as the no-context issue-list baseline.
Note that we evaluate contextual augmentation exclusively under the issue-list setting, rather than primary-issue review, because issue prioritization in the primary-issue setting limits the model's ability to benefit from additional context.
Our best context-enhanced setting (\sit{List-Nbr\textbar Sim}) improves \sit{RefineEM} by 3.76 pp over \sit{List-None}, whereas the same strategy under primary-issue review (\sit{Primary-Nbr\textbar Sim}) yields only 1.16 pp over \sit{Primary-None}.
This narrower margin makes it harder to observe effects of different context strategies.
We further discuss the importance of issue-list generation in \S\ref{sec:ablation}.

For \textbf{RQ2a}, \textbf{RQ2b}, and \textbf{RQ2c}, we introduce \textbf{\sit{List-Nbr}}, \textbf{\sit{List-Sem}}, and \textbf{\sit{List-Sim}}, which use only neighboring, LSP-based semantic, and IR-based similar co-change context as contextual augmentation, respectively, to isolate the individual contribution of each context type.
For \textbf{RQ2d}, we explore how to combine multiple context types for optimal performance. Specifically, we evaluate four combination variants:
\textbf{\sit{List-Nbr\textbar Sem}} (neighboring + semantic),
\textbf{\sit{List-Nbr\textbar Sim}} (neighboring + similar),
\textbf{\sit{List-Sem\textbar Sim}} (semantic + similar),
and \textbf{\sit{List-Nbr\textbar Sem\textbar Sim}} (all three context types).

As a complement to \textbf{RQ2a}, we introduce an additional baseline, \textbf{\sit{List-Random}}, which partitions all files co-modified in the same commit into fixed 20-line blocks and randomly samples 8 blocks as context.
This setting is calibrated to match the context budget of our best-performing augmentation strategy, \sit{List-Nbr\textbar Sim}, which adds about 150 lines of context on average.
We use this baseline to test whether simply adding more context is sufficient.

As a complement to \textbf{RQ2b}, we further ablate the contribution of different LSP query types within \sit{List-Sem}.
Specifically, we introduce two ablated variants:
\textbf{\sit{List-Sem}$_\text{Def}$}, which triggers only \texttt{getDefinition} (together with its companion \texttt{getTypeDefinition}) for each identifier in the modified lines;
and \textbf{\sit{List-Sem}$_\text{Ref}$}, which triggers only \texttt{getReferences}.
Beyond the default identifier-exhaustive query strategy used in \sit{List-Sem}, we also design a change-driven semantic-context strategy, denoted as \textbf{\sit{List-Chg}}, to examine whether a more selective LSP query plan can yield more effective semantic context.
We describe the change-driven query rules in \S\ref{sec:context}.

As a complement to \textbf{RQ2c}, we conduct a sensitivity analysis on two design choices in the IR-based similar co-change context retrieval: (1) block granularity and (2) the similarity filter $\mathcal{F}_{\text{sim}}$. 
For block granularity, our default setting uses $S_{\text{target}}\!=\!20$, $S_{\min}\!=\!10$, $S_{\max}\!=\!30$, $\tau_{\text{merge}}\!=\!10$, and $\lambda_{\text{cut}}\!=\!50$, which is intended to produce medium-sized blocks.
We compare this default medium setting against a \sit{large}-block setting ($S_{\text{target}}\!=\!50$, $S_{\min}\!=\!20$, $S_{\max}\!=\!80$, $\tau_{\text{merge}}\!=\!20$, $\lambda_{\text{cut}}\!=\!200$) and a \sit{small}-block setting ($S_{\text{target}}\!=\!10$, $S_{\min}\!=\!5$, $S_{\max}\!=\!15$, $\tau_{\text{merge}}\!=\!5$, $\lambda_{\text{cut}}\!=\!10$), while keeping all other parameters fixed.
For the similarity filter, we fix the default medium-block setting and vary the filtering threshold in $\mathcal{F}_{\text{sim}}$.
In addition to the default setting ($\mathrm{Score}(\textit{block})\!>\!0$), we further evaluate two stricter variants: $\mathrm{Score}(\textit{block})\!\geq\!0.1$ and $\mathrm{Score}(\textit{block})\!\geq\!0.2$.

\begin{figure}[t]
	\centering
    \setlength{\abovecaptionskip}{3pt}
	\includegraphics[width=0.46\textwidth]{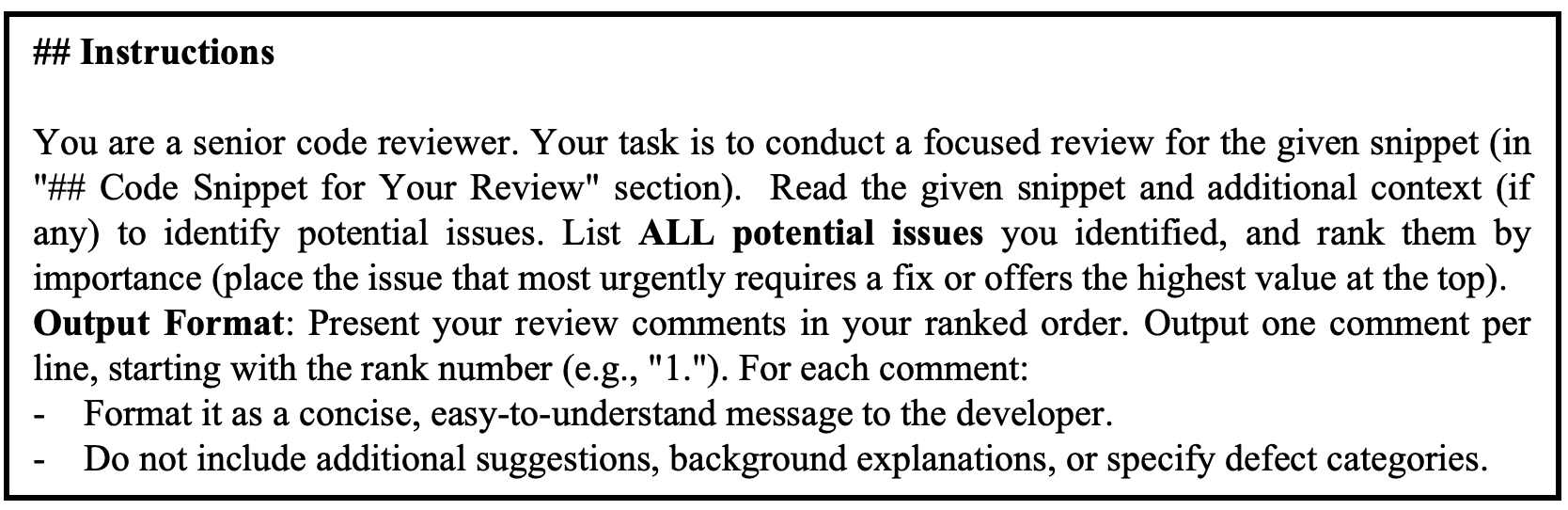}
	\caption{The simplified prompt containing only the task description and output format, without
  the detailed defect taxonomies and review guidelines.}
	\label{fig:prompt_comparison}
\end{figure}

As a complement to \textbf{RQ2d}, to focus purely on the effect of different context sources, we conduct an additional experiment where we replace the original task instruction with a simplified prompt (Figure~\ref{fig:prompt_comparison}) that removes the defect taxonomy and review guidelines.

To answer \textbf{RQ3}, we evaluate the candidate integration strategy described in §\ref{sec:filter}.
We first construct a unified candidate pool by  merging candidates from the best context-enhanced setting and \sit{List-None} (i.e. $\sit{Cands}_{\sit{pool}}\!=\!\sit{Cands}_{\sit{ctx}}\!\oplus\!\sit{Cands}_{\sit{base}}$), and refer to this configuration as \textbf{\sit{List-Merged}}.
This allows us to examine whether integrating candidates generated under different contextual conditions (with and without contextual augmentation), can further improve over the best single configuration identified in RQ2.
We then apply our refinement-guided pruning strategy to $\sit{Cands}_{\sit{pool}}$, referring to this configuration as \textbf{\sit{List-Pruned}}.
To assess whether this strategy is effective at improving candidate-pool quality, we further introduce \textbf{\sit{List-Rerank}}, which applies a LLM-based comment-quality ranking approach proposed by Lu et al.~\cite{lu2025deepcrceval} to the same pool.
Given the reviewed hunk and its candidate comments, this approach first scores each comment along ten quality dimensions (e.g., Readability, Relevance, Explanation Clarity), and then applies chain-of-thought reasoning to produce a final ranking of the candidate list.
We compare \sit{List-Merged}, \sit{List-Pruned}, and \sit{List-Rerank} at top-1, top-3, top-5, and full-list cutoffs.

We use DeepSeek-V3~\cite{deepseekv3}, a cost-efficient model with competitive code-understanding capability, as the underlying model for all review generation settings, with temperature set to 0 to minimize output randomness.
To further mitigate output variability, we run each configuration three times and report the
average results.
For LSP-based program analysis, we use \texttt{gopls}~\cite{gopls} as the Go language server.

\subsection{Dataset}
We build our evaluation dataset from the benchmark released by Li et al. for CodeReviewer~\cite{li2022automating}.
That benchmark covers three related tasks: code change quality estimation, review comment generation (the task we target), and code refinement (whose data we repurpose as described below).
Its review comment generation dataset, however, does not provide repository names or PR IDs, making it impossible to reconstruct the development context required by our research.
We therefore start from the code refinement dataset, which retains this metadata, and reconstruct a review comment generation evaluation set from it.

CodeReviewer covers nine programming languages.
Since our approach relies on LSP-based program analysis, we focus on Go as a representative language.
Go is one of the three most common languages in the CodeReviewer benchmark (after Python and Java), and also a practical choice for reconstructing development context at scale:
Compared with Go, Python’s dynamic typing introduces ambiguity into static symbol resolution.
Java requires resolving project-specific build tools (e.g., Maven or Gradle), which makes environment setup substantially heavier.

\begin{table*}[t]
\caption{\textbf{(RQ1)} Effectiveness of \textbf{Issue-List Review vs. Primary-Issue Review}.}
\label{tab:rq1-results}
\centering
\scriptsize
\begin{tabular}{ccccccc}
\hline
 & \textbf{Avg. Cands} & \textbf{EM Cases} & \textbf{RefineEM} & \textbf{RefineBLEU} & \textbf{ReviewBERT} & \textbf{ReviewBLEU} \\
\hline
\textbf{\sit{Human-Oracle}} & 1.0 & 519.0 & 36.09\% & 84.58 & -- & -- \\
\textbf{\sit{CodeReviewer}} & 1.0 & 216.0 & 15.02\% & 78.80 & 33.96 & 5.62 \\
\textbf{\sit{Primary-None}} ($\triangle$ vs. \sit{CodeReviewer}) & 1.0 (+0.0) & 246.7 (+30.7) & 17.15\% (+2.13\%) & 79.19 (+0.39) & 35.95 (+1.99) & 5.22 (-0.40) \\
\textbf{\sit{List-None}} ($\triangle$ vs. \sit{Primary-None}) & 3.1 (+2.1) & 314.0 (+67.3) & 21.83\% (+4.68\%) & 83.62 (+4.43) & 40.27 (+4.32) & 5.60 (+0.38) \\
\hline
\end{tabular}
\end{table*}

The code refinement test set contains 2,889 comments for Go.
Using the repository names and PR IDs provided in the dataset, we successfully recovered 2,786 corresponding instances; the remaining were excluded due to inaccessible repositories (e.g. deleted, transferred, or made private).
To better match the distribution of the original review comment generation dataset, we retained only hunks with at most 20 lines, yielding a final evaluation set of \textbf{1,438} review instances.
These instances span 54 repositories and 1,282 distinct commit snapshots.
% Each instance forms a complete triad: the pre-change code hunk, the human-written review comment, and the corresponding human-authored code refinement, serving as the ground truth for our subsequent evaluations.

\subsection{Evaluation Metrics}
We evaluate the quality of a generated comment list from two perspectives: review comment quality and induced code refinement quality.

\vspace{1ex}
\textbf{Review Comment Metrics.}
Following Li et al.~\cite{li2022automating}, we measure the textual similarity between a generated comment and the human-written reference using BLEU score~\cite{papineni2002bleu} (\sit{ReviewBLEU}).
We additionally report BERTScore~\cite{DBLP:conf/iclr/ZhangKWWA20} (\sit{ReviewBERT}), computed using \texttt{microsoft/unixcoder-base-nine}~\cite{guo2022unixcoder} as the encoder, which better captures semantic similarity in software engineering text than n-gram overlap alone.
Since our approach produces a ranked comment list rather than a single comment, we assess the \textit{potential} of each list by selecting its best candidate via oracle selection over the generated candidates, where \sit{ReviewBERT} serves as the primary criterion and \sit{ReviewBLEU} as a tie-breaker:
\[
c^* = \arg\max_{c_i \in \sit{Cands}} \bigl(\sit{ReviewBERT}(c_i),\; \sit{ReviewBLEU}(c_i)\bigr).
\]
We report the average \sit{ReviewBLEU} and \sit{ReviewBERT} of $c^*$ across all test cases as the upper-bound performance of the candidate list on review comment quality.

\vspace{1ex}
\textbf{Code Refinement Metrics.}
We further evaluate each candidate by the code change it induces.
Specifically, we feed it into a CodeReviewer model fine-tuned for code refinement to produce a predicted revision, and measure its agreement with the ground-truth revision using exact match (\sit{RefineEM}) and BLEU score (\sit{RefineBLEU}).
To select the optimal candidate under the refinement oracle, we prioritize \sit{RefineEM} and use \sit{RefineBLEU} as a tie-breaker:
  \[
  c^* = \arg\max_{c_i \in \sit{Cands}} \bigl(\sit{RefineEM}(c_i),\;
  \sit{RefineBLEU}(c_i)\bigr),
  \]
We report both \textit{EM Cases} and the \sit{RefineEM} rate.
Here, \textit{EM Cases} denotes the absolute number of test cases for which $c^*$ achieves exact match.
Since our candidate-pruning strategy may leave some cases without any retained candidate, we define: \(\sit{RefineEM}=\textit{EM Cases}/N_{\text{retained}}\), where \(N_{\text{retained}}\) is the number of cases that retain at least one candidate after pruning.
This formulation accounts for the practical value of abstaining from low-value comments.
We also report the average \sit{RefineBLEU} of $c^*$ over these retained cases.
Our primary refinement metrics are \textit{EM Cases} and \sit{RefineEM}, as they directly reflect the extent to which generated comments trigger the same code refinement as the human reviewer.
In addition, we report the average number of candidates per test case to quantify the practical cost of each configuration.

To assess whether the performance difference between two settings is statistically significant, we apply the exact McNemar test~\cite{mcnemar1947note} on per-case binary EM outcomes across the 1,438 instances.
We report Cohen's \(g\)~\cite{cohen2013statistical} as the paired effect size averaged across the three runs.
Following common conventions, we interpret \(g \geq 0.05\), \(g \geq 0.15\), and \(g \geq 0.25\) as small, medium, and large effects, respectively.

%%%%%%%%%%%%%%%%%%%%%%%%%%%%%%%%%%%%%%%% Results %%%%%%%%%%%%%%%%%%%%%%%%%%%%%%%%%%%%%%%%
\section{Results and Discussion}

\subsection{RQ1: Issue-List Generation}
Table~\ref{tab:rq1-results} presents the results under different review paradigms.
Even under the more constrained \sit{Primary-None} setting, our LLM-based approach already outperforms the supervised CodeReviewer baseline, resolving 30.7 more refinement cases and improving across most metrics.
This suggests that LLMs bring strong code understanding capabilities to the review task even without task-specific fine-tuning.
Switching to issue-list review further unlocks the model's ability to surface human-aligned review comments.
Compared to \sit{Primary-None}, \sit{List-None} resolves 67.3 additional cases and significantly improves \sit{RefineEM} by 4.68 pp (McNemar exact $p\!<\!0.05$, Cohen's $g\!=\!0.26$, large effect), with consistent gains across all metrics, while increasing the average candidate count to 3.1.

\begin{table*}[t]
\caption{\textbf{(RQ2)} Effectiveness of \textbf{Contextual Augmentation Strategies}. Avg.\ LOC = average lines of context added per reviewed hunk.}
\label{tab:rq2-results}
\centering
\scriptsize
\begin{tabular}{cccccccc}
\hline
 & \textbf{Avg. LOC} & \textbf{Avg. Cands} & \textbf{EM Cases} & \textbf{RefineEM} & \textbf{RefineBLEU} & \textbf{ReviewBLEU} & \textbf{ReviewBERT} \\
\hline
\textbf{\sit{List-None}} & 0 & 3.1 & 314.0 & 21.83\% & 83.62 & 40.27 & 5.66 \\
\hline
\textbf{\sit{List-Random}} & 149.0 & 3.9 & 331.3 & 23.04\% & 84.34 & 41.10 & 5.69 \\
\textit{($\triangle$ vs.\ \sit{List-None})} & \quad\textit{+149.0} & \quad\textit{+0.8} & \quad\textit{+17.3} & \quad\textit{+1.21} & \quad\textit{+0.73} & \quad\textit{+0.83} & \quad\textit{+0.03} \\
\hline
\textbf{\sit{List-Nbr}} & 66.8 & 3.7 & 339.7 & 23.62\% & 84.32 & 41.11 & 5.85 \\
\textit{($\triangle$ vs.\ \sit{List-None})} & \quad\textit{+66.8} & \quad\textit{+0.6} & \quad\textit{+25.7} & \quad\textit{+1.79} & \quad\textit{+0.71} & \quad\textit{+0.83} & \quad\textit{+0.19} \\
\textbf{\sit{List-Sem}} & 81.4 & 3.7 & 341.0 & 23.71\% & 84.34 & 41.32 & 5.95 \\
\textit{($\triangle$ vs.\ \sit{List-None})} & \quad\textit{+81.4} & \quad\textit{+0.6} & \quad\textit{+27.0} & \quad\textit{+1.88} & \quad\textit{+0.73} & \quad\textit{+1.05} & \quad\textit{+0.29} \\
\textbf{\sit{List-Sim}} & 83.2 & 3.9 & 345.0 & 23.99\% & 84.43 & 40.82 & 5.53 \\
\textit{($\triangle$ vs.\ \sit{List-None})} & \quad\textit{+83.2} & \quad\textit{+0.8} & \quad\textit{+31.0} & \quad\textit{+2.16} & \quad\textit{+0.81} & \quad\textit{+0.55} & \quad\textit{-0.13} \\
\hline
\textbf{\sit{List-Nbr\textbar Sem}} & 134.6 & 3.9 & 358.0 & 24.90\% & 84.56 & 41.38 & 5.84 \\
\textit{($\triangle$ vs.\ \sit{List-Nbr})} & \quad\textit{+67.8} & \quad\textit{+0.2} & \quad\textit{+18.3} & \quad\textit{+1.28} & \quad\textit{+0.24} & \quad\textit{+0.27} & \quad\textit{-0.01} \\
\textit{($\triangle$ vs.\ \sit{List-Sem})} & \quad\textit{+53.2} & \quad\textit{+0.2} & \quad\textit{+17.0} & \quad\textit{+1.18} & \quad\textit{+0.22} & \quad\textit{+0.05} & \quad\textit{-0.11} \\
\textbf{\sit{List-Nbr\textbar Sim}} & 149.0 & 4.1 & \textbf{368.0} & \textbf{25.59\%} & 84.70 & 41.81 & 5.78 \\
\textit{($\triangle$ vs.\ \sit{List-Nbr})} & \quad\textit{+82.2} & \quad\textit{+0.4} & \quad\textit{+28.3} & \quad\textit{+1.97} & \quad\textit{+0.37} & \quad\textit{+0.70} & \quad\textit{-0.07} \\
\textit{($\triangle$ vs.\ \sit{List-Sim})} & \quad\textit{+65.8} & \quad\textit{+0.3} & \quad\textit{+23.0} & \quad\textit{+1.60} & \quad\textit{+0.27} & \quad\textit{+0.99} & \quad\textit{+0.25} \\
\textbf{\sit{List-Sem\textbar Sim}} & 162.7 & 4.0 & 347.7 & 24.18\% & 84.67 & 41.75 & 5.95 \\
\textit{($\triangle$ vs.\ \sit{List-Sem})} & \quad\textit{+81.3} & \quad\textit{+0.3} & \quad\textit{+6.7} & \quad\textit{+0.46} & \quad\textit{+0.33} & \quad\textit{+0.43} & \quad\textit{+0.00} \\
\textit{($\triangle$ vs.\ \sit{List-Sim})} & \quad\textit{+79.5} & \quad\textit{+0.1} & \quad\textit{+2.7} & \quad\textit{+0.19} & \quad\textit{+0.24} & \quad\textit{+0.93} & \quad\textit{+0.42} \\
\hline
\textbf{\sit{List-Nbr\textbar Sem\textbar Sim}} & 215.5 & 4.3 & 357.7 & 24.87\% & 84.86 & 42.07 & 5.96 \\
\textit{($\triangle$ vs.\ \sit{List-Nbr\textbar Sem})} & \quad\textit{+80.9} & \quad\textit{+0.4} & \quad\textit{-0.3} & \quad\textit{-0.03} & \quad\textit{+0.30} & \quad\textit{+0.69} & \quad\textit{+0.12} \\
\textit{($\triangle$ vs.\ \sit{List-Nbr\textbar Sim})} & \quad\textit{+66.5} & \quad\textit{+0.0} & \quad\textit{-10.3} & \quad\textit{-0.72} & \quad\textit{+0.07} & \quad\textit{+0.31} & \quad\textit{+0.21} \\
\textit{($\triangle$ vs.\ \sit{List-Sem\textbar Sim})} & \quad\textit{+52.8} & \quad\textit{+0.3} & \quad\textit{+10.0} & \quad\textit{+0.69} & \quad\textit{+0.19} & \quad\textit{+0.32} & \quad\textit{+0.01} \\
\hline
\end{tabular}
\end{table*}

\begin{figure}[t]
	\centering
    \setlength{\abovecaptionskip}{3pt}
	\includegraphics[width=0.36\textwidth]{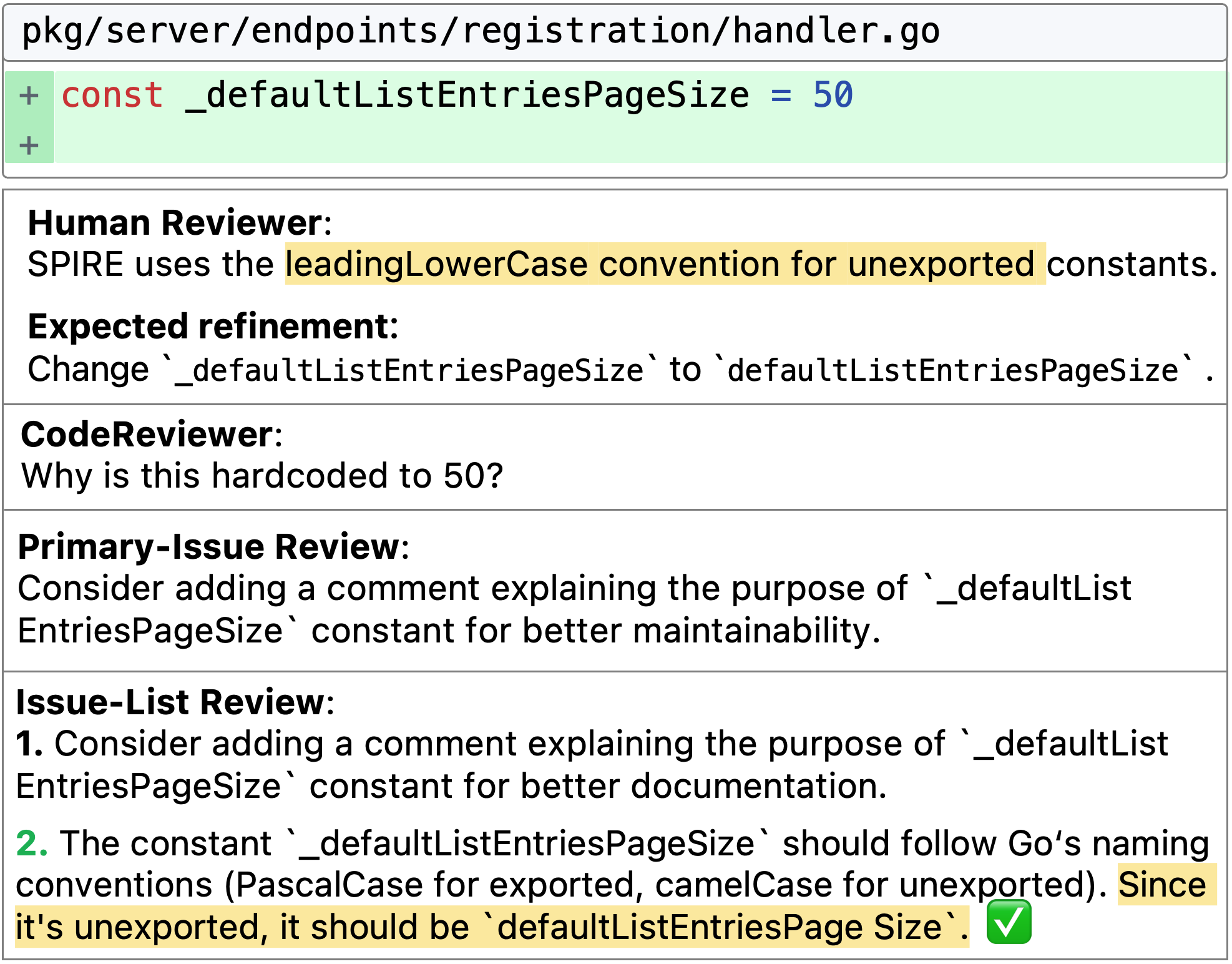}
	\caption{An example illustrating how issue-list review recovers a human-aligned comment (rank~2) that primary-issue review misses.}
	\label{fig:rq1_example}
\end{figure}

Figure~\ref{fig:rq1_example} shows a \href{https://github.com/spiffe/spire/pull/1392#discussion_r380990948}{case} that illustrates why issue-list review recovers comments that primary-issue review misses.
The reviewed change introduces a new unexported constant \texttt{\_defaultListEntriesPageSize}.
The human review points out that the leading underscore violates the Go naming convention for unexported constants, and the code is subsequently revised accordingly.
CodeReviewer questions the hardcoded value (``Why is this hardcoded to 50?''), but without suggesting a concrete fix.
Both primary-issue review and the top-ranked comment of issue-list review recommend adding a comment to explain the purpose of the constant, a valid and actionable suggestion, but not the one the human reviewer prioritized.
However, the rank-2 comment of issue-list review identifies the naming convention violation, consistent with the human reviewer's comment.
This reveals that primary-issue review may miss the correct suggestion not because the issue cannot be detected, but because it is overshadowed by other seemingly more urgent suggestions.
In contrast, issue-list review can mitigate this problem by surfacing lower-ranked comments.

\begin{rqanswerbox}
\textbf{Answer to RQ1:}
Issue-list review (\sit{List-None}) already significantly improves \sit{RefineEM} to 21.83\% over primary-issue review (17.15\% \sit{RefineEM}) by \textbf{4.68} pp, while increasing the average candidate count from 1.0 to 3.1.
\end{rqanswerbox}

\subsection{RQ2: Contextual Augmentation}

\vspace{1ex}
\subsubsection{RQ2.a Neighboring Context}
Table~\ref{tab:rq2-results} presents the results under different contextual augmentation strategies.
The random baseline \sit{List-Random} already improves over \sit{List-None}, resolving 17.3 additional cases (+1.21 pp \sit{RefineEM}).
However, all designed retrieval strategies outperform it, indicating that review effectiveness depends not only on the amount of added context but also on how that context is selected.
Specifically, adding function-level neighboring context (\sit{List-Nbr}) improves over \sit{List-None}, resolving 25.7 additional cases and improving \sit{RefineEM} by 1.79 pp ($p\!<\!0.05$, Cohen's $g\!=\!0.12$, small effect).

% Building on \sit{List-Nbr}, both \sit{List-Sem} and \sit{List-Sim} extend context beyond spatial locality, and both yield further gains:
% \sit{List-Sem} resolves 18.3 additional cases (+1.28\% \sit{RefineEM}), while \sit{List-Sim} resolves 28.3 additional cases (+1.97\% \sit{RefineEM}), making \sit{List-Sim} the more effective augmentation strategy.
% However, combining both in \sit{List-All} does not yield additive improvements: it resolves 10.3 \emph{fewer} cases than \sit{List-Sim} alone, with a 0.72\% drop in \sit{RefineEM}, despite adding more context (216.8 vs.\ 149.0 Avg.\ LOC).
% To better understand how each context type contributes to review effectiveness, and why their combination underperforms, we conduct the following additional analysis.

\vspace{1ex}
\subsubsection{RQ2.b Semantic Context}
\sit{List-Sem} achieves a slightly higher improvement than \sit{List-Nbr}, resolving 27.0 additional cases over \sit{List-None} and increasing \sit{RefineEM} by 1.88 pp ($p\!<\!0.05$ in 2/3 runs, Cohen's $g\!=\!0.12$, small effect).
Table~\ref{tab:rq2-sem-ablation} reports the ablation results for different LSP query types and the change-driven query strategy.
The results show that our default identifier-exhaustive query strategy achieves the best performance among all semantic context variants we evaluated.
The detailed analysis is as follows.

\vspace{1ex}
\textit{Ablation of LSP Query Types.}
We find that definition-related and reference-related context contribute comparable improvements, with \sit{List-Sem}$_\text{Def}$ recovering 15.7 additional cases (+1.09 pp \sit{RefineEM}) and \sit{List-Sem}$_\text{Ref}$ recovering 19.3 (+1.35 pp).
Nevertheless, \sit{List-Sem}$_\text{Def}$ achieves this with less added context (26.6 lines), compared to 57.2 lines for \sit{List-Sem}$_\text{Ref}$, making definition-related context the more cost-efficient choice.
% Moreover, when combined in \sit{List-Sem}, the two query types exhibit a synergistic effect: \sit{List-Sem} resolves 18.3 additional cases, more than the sum of their individual contributions.
Figure~\ref{fig:denifition_example} provides an example of how definition context contributes to review effectiveness.
In this \href{https://github.com/open-telemetry/opentelemetry-go/pull/2275#discussion_r722029637}{case} from \texttt{open-telemetry/opentelemetry-go}, the change adds a validity check before calling \texttt{s.addLink(l)}.
The reviewer suggests moving this check into \texttt{addLink} itself to centralize the validation logic.
When definition context is available, \texttt{getDefinition} on \texttt{addLink} retrieves its implementation, helping the model infer that the check is better placed inside the function.
Without access to such implementation details, models (e.g., \sit{List-Sem}$_\text{Ref}$) tend to miss this suggestion entirely.
% Beyond such direct contributions, definition context can also help more implicitly.
% For example, in a \href{https://github.com/fleetdm/fleet/pull/1441#discussion_r674287109}{case} from \texttt{fleetdm/fleet}, the change modifies an argument to \texttt{addConfigString}, replacing a hardcoded path with a \texttt{fmt.Sprintf} expression, while the reviewer suggests using \texttt{filepath.Join} instead.
% Both \sit{List-Sem}$_\text{Def}$ and \sit{List-Sem} recover this suggestion.
% A likely reason is that the definition of \texttt{addConfigString(key, defVal, usage string)} makes clear that the argument is intended to be a plain path string, prompting the model to consider a more idiomatic string-construction method.

% \begin{table}[t]
% \caption{\textbf{(RQ2)} Effectiveness of semantic context under different LSP query strategies. $\triangle$ relative to \sit{List-Nbr}.}
% \label{tab:rq2-sem-ablation}
% \centering
% \footnotesize
% \setlength{\tabcolsep}{4pt}
% \renewcommand{\arraystretch}{1.0}
% \begin{tabular}{cccc}
% \hline
% \textbf{Setting} & \textbf{Avg. LOC} & \textbf{EM Cases} & \textbf{RefineEM} \\
% \hline
% \textbf{\sit{List-Nbr}} & 66.8 & 339.7 & 23.62\% \\
% \hline
% \textbf{\sit{List-Sem}$_\text{Def}$} & 88.4 (+21.7) & 348.0 (+8.3) & 24.20\% (+0.58\%) \\
% \textbf{\sit{List-Sem}$_\text{Ref}$} & 113.2 (+46.3) & 344.0 (+4.3) & 23.92\% (+0.30\%) \\
% \textbf{\sit{List-Sem}} & 134.6 (+67.8) & 358.0 (+18.3) & 24.90\% (+1.28\%) \\
% \hline
% \textbf{\sit{List-Sem}$_\text{Chg}$} & 88.5 (+21.7) & 347.3 (+7.7) & 24.15\% (+0.53\%) \\
% \hline
% \end{tabular}
% \end{table}

\begin{table}[t]
\caption{\textbf{(RQ2)} Effectiveness of semantic context under different LSP query strategies. $\triangle$ relative to \sit{List-None}.}
\label{tab:rq2-sem-ablation}
\centering
\scriptsize
\begin{tabular}{cccc}
\hline
\textbf{Setting} & \textbf{Avg. LOC} & \textbf{EM Cases} & \textbf{RefineEM} \\
\hline
\textbf{\sit{List-None}} & 0 & 314.0 & 21.83\% \\
\hline
\textbf{\sit{List-Sem}$_\text{Def}$} & 26.6 (+26.6) & 329.7 (+15.7) & 22.93\% (+1.09) \\
\textbf{\sit{List-Sem}$_\text{Ref}$} & 57.2 (+57.2) & 333.3 (+19.3) & 23.18\% (+1.35) \\
\textbf{\sit{List-Sem}} & 81.4 (+81.4) & 341.0 (+27.0) & 23.71\% (+1.88) \\
\hline
\textbf{\sit{List-Sem}$_\text{Chg}$} & 26.5 (+26.5) & 339.0 (+25.0) & 23.57\% (+1.74) \\
\hline
\end{tabular}
\end{table}

\begin{table}[t]
\caption{\textbf{Change-driven Query Rules} for LSP-Based Semantic Context.}
\label{tab:change-driven-rules}
\centering
\scriptsize
\setlength{\tabcolsep}{2pt}
\renewcommand{\arraystretch}{0.95}
\begin{tabular}{p{0.06\columnwidth} >{\raggedright\arraybackslash}p{0.88\columnwidth}}
\hline
\textbf{Rule} & \multicolumn{1}{c}{\textbf{Trigger Pattern $\rightarrow$ Issued Query}} \\
\hline
\multicolumn{2}{l}{\textbf{Type-level rules:}} \\
R1  & A new import is added \newline $\rightarrow$ \texttt{getReferences} on the imported package/module. \\
R2  & A new type reference is added $\rightarrow$ \texttt{getDefinition} on that type. \\
R3  & A new type definition is added $\rightarrow$ \texttt{getReferences} on that type. \\
R4  & The type's inheritance or implementation relationships are changed
\newline$\rightarrow$ \texttt{getDefinition} on the changed related type
\newline and \texttt{getReferences} on the current type. \\
R5  & A type is renamed or deleted $\rightarrow$ \texttt{getSearch} on the old type name. \\
\hline
\multicolumn{2}{l}{\textbf{Function-level rules:}} \\
R6  & A new function call is added $\rightarrow$ \texttt{getDefinition} on the called function. \\
R7  & A function call's arguments are modified
      \newline $\rightarrow$ \texttt{getDefinition} on the called function. \\
R8  & A new function definition is added $\rightarrow$ \texttt{getReferences} on that function. \\
R9 & A function's signature is modified $\rightarrow$ \texttt{getReferences} on that function. \\
\hline
\multicolumn{2}{l}{\textbf{Variable-level rules:}} \\
R10 & A new variable reference is added $\rightarrow$ \texttt{getDefinition} on that identifier. \\
R11 & A new variable declaration is added $\rightarrow$ \texttt{getReferences} on that identifier. \\
R12 & The value of a variable is modified
    \newline $\rightarrow$ both \texttt{getDefinition} and \texttt{getReferences} on that identifier. \\
R13 & A string literal is assigned to a variable $\rightarrow$ \texttt{getSearch} on that literal. \\
\hline
\end{tabular}
\end{table}

\vspace{1ex}
\textit{Change-Driven Query Rules For Semantic Context.}
Compared to the default identifier-exhaustive query strategy, we further design a change-driven query strategy to examine whether a more selective query plan can yield more effective semantic context.
Under this strategy, we use tree-sitter to parse both the pre-change and post-change versions of the reviewed file and identify the specific identifiers introduced or modified by the hunk.
Inspired by prior work on change types~\cite{fluri2006classifying,fluri2007change}, we design a set of 13 change-driven rules to  map specific change patterns to targeted LSP queries (Table~\ref{tab:change-driven-rules}).
These rules are organized around three levels of program elements that a change may involve:
\textit{type-level} rules for changes involving imports, classes, and interfaces;
\textit{function-level} rules for changes involving function definitions, calls, and signatures;
and \textit{variable-level} rules for changes involving parameters, fields, properties, and constants.
When a hunk matches one or more rules, the corresponding LSP queries are executed to retrieve the semantic context.

As shown in Table~\ref{tab:rq2-sem-ablation}, \sit{List-Sem}$_\text{Chg}$ does not outperform \sit{List-Sem} (339.0 vs.\ 341.0 EM cases).
This suggests that focusing queries on changed identifiers alone is insufficient for code review, as context beyond the immediate change is often equally important.
For example, for the Figure~\ref{fig:denifition_example} case, a change-driven strategy would only query for the newly introduced identifiers, \texttt{SpanContext} and \texttt{IsValid}, and would miss the definition of \texttt{addLink}, which is the key context to identify the validation placement issue.
This confirms that our default identifier-exhaustive strategy (i.e. querying all identifiers in modified lines) is more effective for retrieving semantic context.

\vspace{1ex}
\subsubsection{RQ2.c Similar Co-Change Context}
Table~\ref{tab:rq2-results} shows that similar co-change context yields the largest gain over \sit{List-None} among the three single-strategy variants, resolving 31.0 additional cases and improving \sit{RefineEM} by 2.16 pp ($p\!<\!0.05$, Cohen's $g\!=\!0.14$, small effect).

\begin{table}[t]
\caption{\textbf{(RQ2)} Effectiveness of similar context under different block granularities and similarity filter thresholds. $\triangle$ is relative to \sit{List-None}. $\dagger$ = default.}
\label{tab:rq2-sim-ablation}
\centering
\scriptsize
\begin{adjustbox}{max width=0.48\textwidth}
\begin{tabular}{cccc}
\hline
\textbf{Setting} & \textbf{Avg. LOC} & \textbf{EM Cases} & \textbf{RefineEM} \\
\hline
\textbf{\sit{List-None}} & 0 & 314.0 & 21.83\% \\
\hline
\multicolumn{4}{l}{\textit{Block Granularity:}} \\
\hline
\textbf{\sit{List-Sim} (Large)} & 150.9 (+150.9) & 338.7 (+24.7) & 23.55\% (+1.72) \\
\textbf{\sit{List-Sim} (Medium)$^\dagger$} & 83.2 (+83.2) & \textbf{345.0} (+31.0) & \textbf{23.99\%} (+2.16) \\
\textbf{\sit{List-Sim} (Small)} & 45.5 (+45.5) & 336.3 (+22.3) & 23.39\% (+1.55) \\
\hline
\multicolumn{4}{l}{\textit{Similarity Filter ($\mathcal{F}_{\text{sim}}$):}} \\
\hline
\textbf{\sit{List-Sim} ($> 0$)$^\dagger$} & 83.2 (+83.2) & \textbf{345.0} (+31.0) & \textbf{23.99\%} (+2.16) \\
\textbf{\sit{List-Sim} ($\geq 0.1$)} & 82.3 (+82.3) & 342.3 (+28.3) & 23.80\% (+1.97) \\
\textbf{\sit{List-Sim} ($\geq 0.2$)} & 73.9 (+73.9) & 334.7 (+20.7) & 23.27\% (+1.44) \\
\hline
\end{tabular}
\end{adjustbox}
\end{table}

\begin{table}[t]
\caption{\textbf{(RQ2)} Effect of contextual augmentation under different task prompts ($\dagger$ = default).}
\label{tab:rq2-prompt-ablation}
\centering
\scriptsize
\setlength{\tabcolsep}{2pt}
\begin{adjustbox}{max width=0.48\textwidth}
\begin{tabular}{ccccccc}
\hline
& \multicolumn{3}{c}{\textbf{Detail Prompt}$^\dagger$} & \multicolumn{3}{c}{\textbf{Simple Prompt}} \\
\hline
 & \makecell{\textbf{Avg.}\\\textbf{Cands}} & \makecell{\textbf{EM}\\\textbf{Cases}} & \makecell{\textbf{RefineEM}} & \makecell{\textbf{Avg.}\\\textbf{Cands}} & \makecell{\textbf{EM}\\\textbf{Cases}} & \makecell{\textbf{RefineEM}} \\
\hline
\textbf{\sit{List-None}} & 3.1 & 314.0 & 21.83\% & 2.6 & 322.3 & 22.41\% \\
\hline
\textbf{\sit{List-Nbr}} & 3.7 & 339.7 & 23.62\% & 3.0 & 343.7 & 23.90\% \\
\textit{($\triangle$ vs.\ \sit{List-None})} & \quad\textit{+0.6} & \quad\textit{+25.7} & \quad\textit{+1.79\%} & \quad\textit{+0.4} & \quad\textit{+21.3} & \quad\textit{+1.48} \\
\textbf{\sit{List-Sem}} & 3.7 & 341.0 & 23.71\% & 2.9 & 321.0 & 22.34\% \\
\textit{($\triangle$ vs.\ \sit{List-None})} & \quad\textit{+0.6} & \quad\textit{+27.0} & \quad\textit{+1.88\%} & \quad\textit{+0.3} & \quad\textit{-1.3} & \quad\textit{-0.07} \\
\textbf{\sit{List-Sim}} & 3.9 & 345.0 & 23.99\% & 3.0 & 318.0 & 22.14\% \\
\textit{($\triangle$ vs.\ \sit{List-None})} & \quad\textit{+0.8} & \quad\textit{+31.0} & \quad\textit{+2.16\%} & \quad\textit{+0.4} & \quad\textit{-4.3} & \quad\textit{-0.27} \\
\hline
\textbf{\sit{List-Nbr\textbar Sem}} & 3.9 & 358.0 & 24.90\% & 3.3 & 349.3 & 24.29\% \\
\textit{($\triangle$ vs.\ \sit{List-Nbr})} & \quad\textit{+0.2} & \quad\textit{+18.3} & \quad\textit{+1.28\%} & \quad\textit{+0.2} & \quad\textit{+5.7} & \quad\textit{+0.40} \\
\textit{($\triangle$ vs.\ \sit{List-Sem})} & \quad\textit{+0.2} & \quad\textit{+17.0} & \quad\textit{+1.18\%} & \quad\textit{+0.3} & \quad\textit{+28.3} & \quad\textit{+1.95} \\
\textbf{\sit{List-Nbr\textbar Sim}} & 4.1 & \textbf{368.0} & \textbf{25.59\%} & 3.4 & 350.7 & 24.39\% \\
\textit{($\triangle$ vs.\ \sit{List-Nbr})} & \quad\textit{+0.4} & \quad\textit{+28.3} & \quad\textit{+1.97\%} & \quad\textit{+0.3} & \quad\textit{+7.0} & \quad\textit{+0.49} \\
\textit{($\triangle$ vs.\ \sit{List-Sim})} & \quad\textit{+0.3} & \quad\textit{+23.0} & \quad\textit{+1.60\%} & \quad\textit{+0.3} & \quad\textit{+32.7} & \quad\textit{+2.25} \\
\textbf{\sit{List-Sem\textbar Sim}} & 4.0 & 347.7 & 24.18\% & 3.2 & 326.0 & 22.65\% \\
\textit{($\triangle$ vs.\ \sit{List-Sem})} & \quad\textit{+0.3} & \quad\textit{+6.7} & \quad\textit{+0.46\%} & \quad\textit{+0.3} & \quad\textit{+5.0} & \quad\textit{+0.31} \\
\textit{($\triangle$ vs.\ \sit{List-Sim})} & \quad\textit{+0.1} & \quad\textit{+2.7} & \quad\textit{+0.19\%} & \quad\textit{+0.2} & \quad\textit{+8.0} & \quad\textit{+0.51} \\
\hline
\textbf{\sit{List-Nbr\textbar Sem\textbar Sim}} & 4.3 & 357.7 & 24.87\% & 3.5 & \textbf{355.0} & \textbf{24.69\%} \\
\textit{($\triangle$ vs.\ \sit{List-Nbr\textbar Sem})} & \quad\textit{+0.4} & \quad\textit{-0.3} & \quad\textit{-0.03\%} & \quad\textit{+0.3} & \quad\textit{+5.7} & \quad\textit{+0.39} \\
\textit{($\triangle$ vs.\ \sit{List-Nbr\textbar Sim})} & \quad\textit{+0.0} & \quad\textit{-10.3} & \quad\textit{-0.72\%} & \quad\textit{+0.0} & \quad\textit{+4.3} & \quad\textit{+0.30} \\
\textit{($\triangle$ vs.\ \sit{List-Sem\textbar Sim})} & \quad\textit{+0.3} & \quad\textit{+10.0} & \quad\textit{+0.69\%} & \quad\textit{+0.3} & \quad\textit{+29.0} & \quad\textit{+2.04} \\
\hline
\end{tabular}
\end{adjustbox}
\end{table}

\looseness=-1
Table~\ref{tab:rq2-sim-ablation} further reports the sensitivity analysis results for different block granularities and similarity filter thresholds.
For block granularity, the medium setting achieves the best performance (345.0 EM cases), outperforming both the large-block (338.7) and small-block (336.3) variants.
This suggests that overly coarse blocks may introduce more irrelevant content, making the similar patterns less clear as references, while overly fine partitioning may break such patterns into incomplete fragments.
Compared with these two settings, the medium setting appears to be a better design choice.
For the similarity filter, tightening $\mathcal{F}_{\text{sim}}$ from $\mathrm{Score}(\textit{block})\!>\!0$ to $\mathrm{Score}(\textit{block}) \geq\!0.1$ causes only a minor drop in performance, consistent with the fact that this stricter threshold affects only 59 cases.
Nevertheless, the overall trend remains downward as the threshold becomes more restrictive.
This suggests that, beyond the highly similar blocks that provide the primary pattern cues, moderately similar blocks can still contribute useful supplementary context.
Stricter filtering therefore removes potentially helpful context while yielding limited benefit.
Overall, our default settings ($S_{\text{target}}\!=\!20$, $\mathcal{F}_{\text{sim}}\!: \mathrm{Score}(\textit{block})\!>\!0$) remain a reasonable and robust choice.

\vspace{1ex}
\subsubsection{RQ2.d Combining Context}
When combining context types, neighboring context consistently contributes positively: every combination that includes neighboring context outperforms its non-neighbor counterpart.
However, combining semantic and similar co-change context does not yield additive improvements.
\sit{List-Nbr\textbar Sem\textbar Sim} even underperforms \sit{List-Nbr\textbar Sim} by 10.3 EM cases and 0.72 pp in \sit{RefineEM}, despite adding more context.
Overall, our best-performing contextual augmentation strategy is \sit{List-Nbr\textbar Sim}, which pairs neighboring context with the strongest single strategy (similar co-change context), and achieves +54.0 EM cases and improves \sit{RefineEM} by +3.76 pp over \sit{List-None} ($p\!<\!0.05$, Cohen's $g\!=\!0.22$, medium effect).

\begin{figure}[t]
	\centering
    \setlength{\abovecaptionskip}{3pt}
	\includegraphics[width=0.39\textwidth]{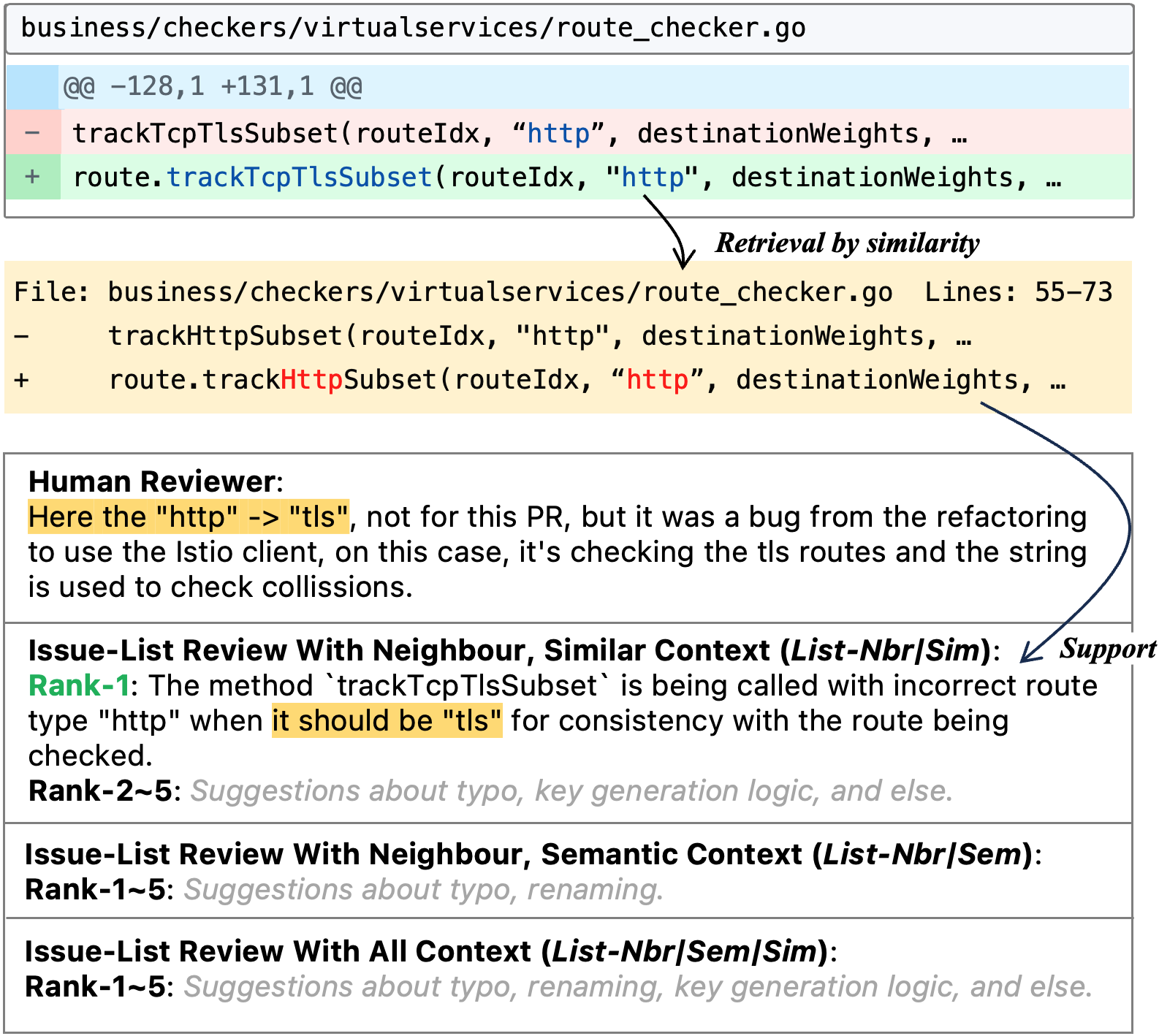}
	\caption{An example of how additional context causes \sit{List-Nbr\textbar Sem\textbar Sim} to miss an otherwise recoverable human-aligned issue.}
	\label{fig:combination_fail_example_1}
\end{figure}

Considering the detailed task prompt may itself interact with context, potentially masking or amplifying their combined effect, we replicate the context combination experiments under a simplified prompt to better isolate the interactions among context types.
The results are shown in Table~\ref{tab:rq2-prompt-ablation}.
Under the simplified prompt, the three context types show complementary effects: \sit{List-Nbr\textbar Sem\textbar Sim} becomes the best-performing setting, whereas under the default prompt it does not.
However, this effect remains limited in magnitude.
The overall best result under the simplified prompt (355.0 EM cases) still falls short of \sit{List-Sim} under the default prompt (368.0 EM cases).
This suggests that, when the model's effective input budget is limited, it is better allocated to strong task guidance and an efficient context setting, rather than distributed across a simplified instruction and multiple context signals.

Additionally, we find that neighboring context plays a particularly important role under the simplified prompt.
Although the best simplified-prompt setting improves by 32.7 EM cases over \sit{List-None},  context alone: \sit{List-Nbr} improves over \sit{List-None} by 21.3 EM cases, accounting for most of the total gain of 32.7 EM cases achieved by \sit{List-Nbr\textbar Sem\textbar Sim}.
By contrast, \sit{List-Sem}, \sit{List-Sim}, and \sit{List-Sem\textbar Sim}, all of which exclude neighboring context, show only limited gains and can even slightly underperform \sit{List-None} ($-1.3$, $-4.3$, and $+3.7$ EM cases, respectively).
This suggests that, without detailed review guidance, the model first needs to understand the local code logic through neighboring  context before it can effectively leverage semantic and similar context.

More broadly, Table~\ref{tab:rq2-prompt-ablation} highlights the importance of aligning prompt design with context complexity.
For lighter context settings such as \sit{List-None} and \sit{List-Nbr}, a simpler prompt can even perform slightly better (\textit{vs.}\ the detailed prompt: $+8.3$ and $+4.0$ EM cases, respectively).
By contrast, richer and more diverse context benefits more from stronger task guidance.
The consistently higher \textit{Avg.\ Cands} under the detailed prompt across all settings further confirms that explicitly defining defect taxonomies and review guidelines is important for eliciting a thorough review.

To better understand why additional context can sometimes hurt rather than help in LLM-based code review, we examine two representative cases.
First, additional context can redirect the model's attention and cause it to miss a previously recoverable issue.
Figure~\ref{fig:combination_fail_example_1} shows a \href{https://github.com/kiali/kiali/pull/4520#discussion_r759161084}{case} from \texttt{kiali/kiali}, where the reviewed change only rewrites a call from \texttt{trackTcpTlsSubset()} to \texttt{route.trackTcpTlsSubset()}.
However, the human reviewer points out a latent semantic bug: in this TLS branch, the argument \texttt{"http"} should be \texttt{"tls"}.
\sit{List-Nbr\textbar Sim} successfully surfaces this issue, likely because the retrieved similar change shows a parallel pattern in which \texttt{trackHttpSubset} is paired with \texttt{"http"}, making the inconsistency easier to recognize by contrast.
However, when \sit{List-Nbr\textbar Sem\textbar Sim} additionally incorporates semantic context, the model no longer reports this human-aligned issue.

\begin{figure}[t]
	\centering
    \setlength{\abovecaptionskip}{3pt}
	\includegraphics[width=0.39\textwidth]{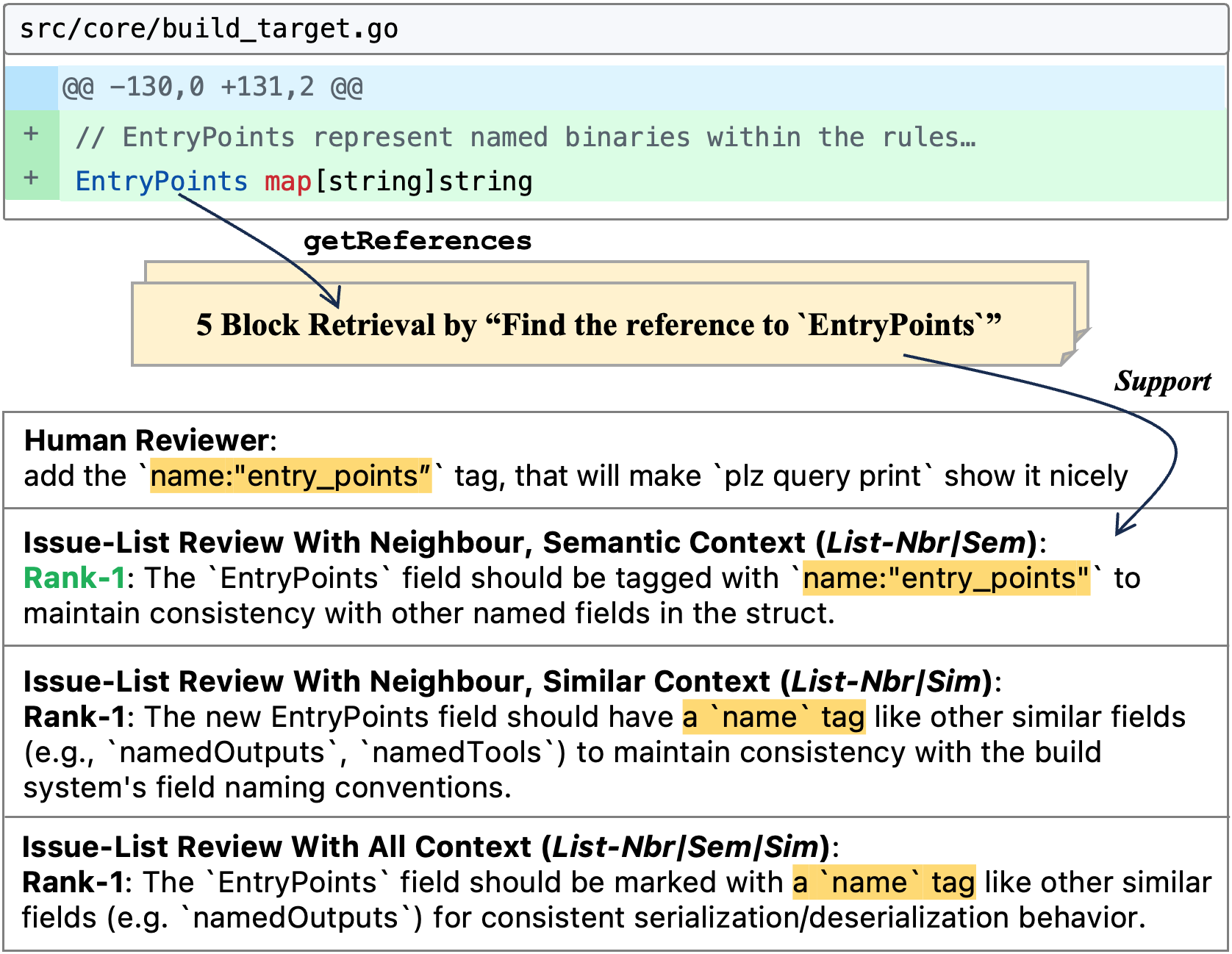}
	\caption{An example of how excessive context causes \sit{List-Nbr\textbar Sem\textbar Sim} to produce a less specific suggestion than \sit{List-Nbr\textbar Sem}.}
	\label{fig:combination_fail_example_2}
\end{figure}

\begin{table*}[t]
\caption{\textbf{(RQ3)} Effectiveness of \textbf{Candidate Integration Strategies}. $\triangle$ values are calculated against their respective baselines at the same cutoff.
\sit{List-Pruned} is evaluated on the 1,408 cases for which at least one candidate is retained.}
\label{tab:rq3-results}
\centering
\scriptsize
\setlength{\tabcolsep}{5pt}
\begin{tabular}{cccccccc}
\hline
\textbf{Setting} & \textbf{Cutoff} & \textbf{Avg. Cands} & \textbf{EM Cases} & \textbf{RefineEM} & \textbf{RefineBLEU} & \textbf{ReviewBLEU} & \textbf{ReviewBERT} \\
\hline
\textbf{\sit{List-Nbr\textbar Sim}} & @All & 4.1 & 368.0 & 25.59\% & 84.70 & 41.81 & 5.78 \\
\hline
\multirow{4}{*}{\makecell{\textbf{\sit{List-Merged}} \\ ($\triangle$ vs.\ \sit{List-Sim})}}
 & @1  & 1.0 & 272.7 & 18.96\% & 80.25 & 34.85 & 5.02 \\
 & @3  & 3.0 & 349.3 & 24.29\% & 84.01 & 40.63 & 5.65 \\
 & @5  & 4.9 & 384.0 & 26.70\% & 85.02 & 42.78 & 6.00 \\
 & @All & 7.2 (+3.1) & 402.7 (+34.7) & 28.00\% (+2.41) & 85.65 (+0.96) & 44.36 (+2.55) & 6.31 (+0.53) \\
\hline
\multirow{4}{*}{\makecell{\textbf{\sit{List-Pruned}} \\ ($\triangle$ vs.\ \sit{List-Merged})}}
 & @1  & 1.0 (+0.0) & 291.3 (+18.7) & 20.69\% (+1.73) & 79.34 (-0.91) & 34.90 (+0.06) & 4.98 (-0.04) \\
 & @3  & 2.4 (-0.6) & 381.7 (+32.3) & 27.11\% (+2.82) & 83.87 (-0.14) & 39.36 (-1.27) & 5.51 (-0.15) \\
 & @5  & 3.1 (-1.8) & 401.0 (+17.0) & 28.48\% (+1.78) & 84.56 (-0.46) & 40.17 (-2.61) & 5.65 (-0.35) \\
 & @All & 3.4 (-3.9) & 402.7 (+0.0) & 28.60\% (+0.60) & 84.72 (-0.94) & 40.40 (-3.97) & 5.69 (-0.62) \\
\hline
\multirow{4}{*}{\makecell{\textbf{\sit{List-Rerank}} \\ ($\triangle$ vs.\ \sit{List-Merged})}}
 & @1  & 1.0 (+0.0) & 259.3 (-13.3) & 18.03\% (-0.93) & 79.42 (-0.83) & 35.04 (+0.20) & 5.02 (+0.00) \\
 & @3  & 3.0 (+0.0) & 336.3 (-13.0) & 23.39\% (-0.90) & 83.45 (-0.56) & 40.61 (-0.02) & 5.81 (+0.16) \\
 & @5  & 4.9 (+0.0) & 376.0 (-8.0)  & 26.15\% (-0.55) & 84.83 (-0.19) & 42.90 (+0.13) & 6.12 (+0.12) \\
 & @All & 7.2 (+0.0) & 402.7 (+0.0) & 28.00\% (+0.00) & 85.65 (+0.00) & 44.36 (+0.00) & 6.31 (+0.00) \\
\hline
\end{tabular}
\end{table*}

Second, excessive context can make the model's suggestion less specific.
Figure~\ref{fig:combination_fail_example_2} shows a \href{https://github.com/thought-machine/please/pull/1243#discussion_r496932095}{case} from \texttt{thought-machine/ please}, where the human reviewer, \sit{List-Nbr\textbar Sem}, and \sit{List-Nbr\textbar Sim} all identify the same high-level issue: the field should have a \texttt{name} tag.
However, among the LLM-based settings, only \sit{List-Nbr\textbar Sem} produces the specific fix \texttt{name:"entry\_points"}, which is necessary for exact-match refinement.
We attribute this to the reference-based semantic context, which repeatedly shows how \texttt{EntryPoints} is consumed in downstream logic, making its intended external interface concrete.
Under \sit{List-Nbr\textbar Sem\textbar Sim}, although the same references are present, the additional context appears to dilute this signal, and the model falls back to the generic suggestion.

Taken together, these cases suggest that contextual augmentation for LLM-based code review is not merely a retrieval problem in which higher contextual recall necessarily leads to better review quality.
Even when useful context is successfully retrieved, the model may still fail to prioritize or exploit it effectively.
Therefore, beyond retrieval, future approaches should also explore better ways to select, organize, and present context, so that useful evidence is reinforced rather than diluted by competing information.

\begin{rqanswerbox}
\textbf{Answer to RQ2:}
Neighboring context, semantic context, and similar co-change context each improve \sit{RefineEM} over the no-context issue-list setting (\sit{List-None}) by \textbf{1.79} pp, \textbf{1.88} pp, and \textbf{2.16} pp, respectively (RQ2a--RQ2c).
However, combining all three context (\sit{List-Nbr\textbar Sem\textbar Sim}) does \textbf{not} achieve the best performance.
Overall, \sit{List-Nbr\textbar Sim}, which combines neighboring context with IR-based similar co-change context, is the best-performing contextual augmentation strategy, achieving an overall \sit{RefineEM} improvement of \textbf{3.76} pp over \sit{List-None} (RQ2d).
\end{rqanswerbox}

\subsection{RQ3: Candidate Integration}
As shown in Table~\ref{tab:rq3-results}, merging candidates from \sit{List-Nbr\textbar Sim} and \sit{List-None} into a unified pool (\sit{List-Merged}) further improves review effectiveness, resolving 34.7 more cases and improving \sit{RefineEM} by 2.41 pp over \sit{List-Nbr\textbar Sim} ($p\!<\!0.05$, Cohen's $g\!=\!0.50$, large effect)\footnote{Since \sit{List-Merged} is a superset of \sit{List-Nbr\textbar Sim}, no case solved by the latter can be unsolved by the former, so all discordant pairs favour \sit{List-Merged} and $g$ reaches its theoretical maximum of 0.50 by construction.}.
In total, \sit{List-Merged} reaches 402.7 EM cases (28.00\% \sit{RefineEM}), an overall improvement of 10.85 pp over \sit{Primary-None} ($p\!<\!0.05$, Cohen's $g\!=\!0.43$, large effect).

\begin{figure}[t]
	\centering
    \setlength{\abovecaptionskip}{3pt}
	\includegraphics[width=0.40\textwidth]{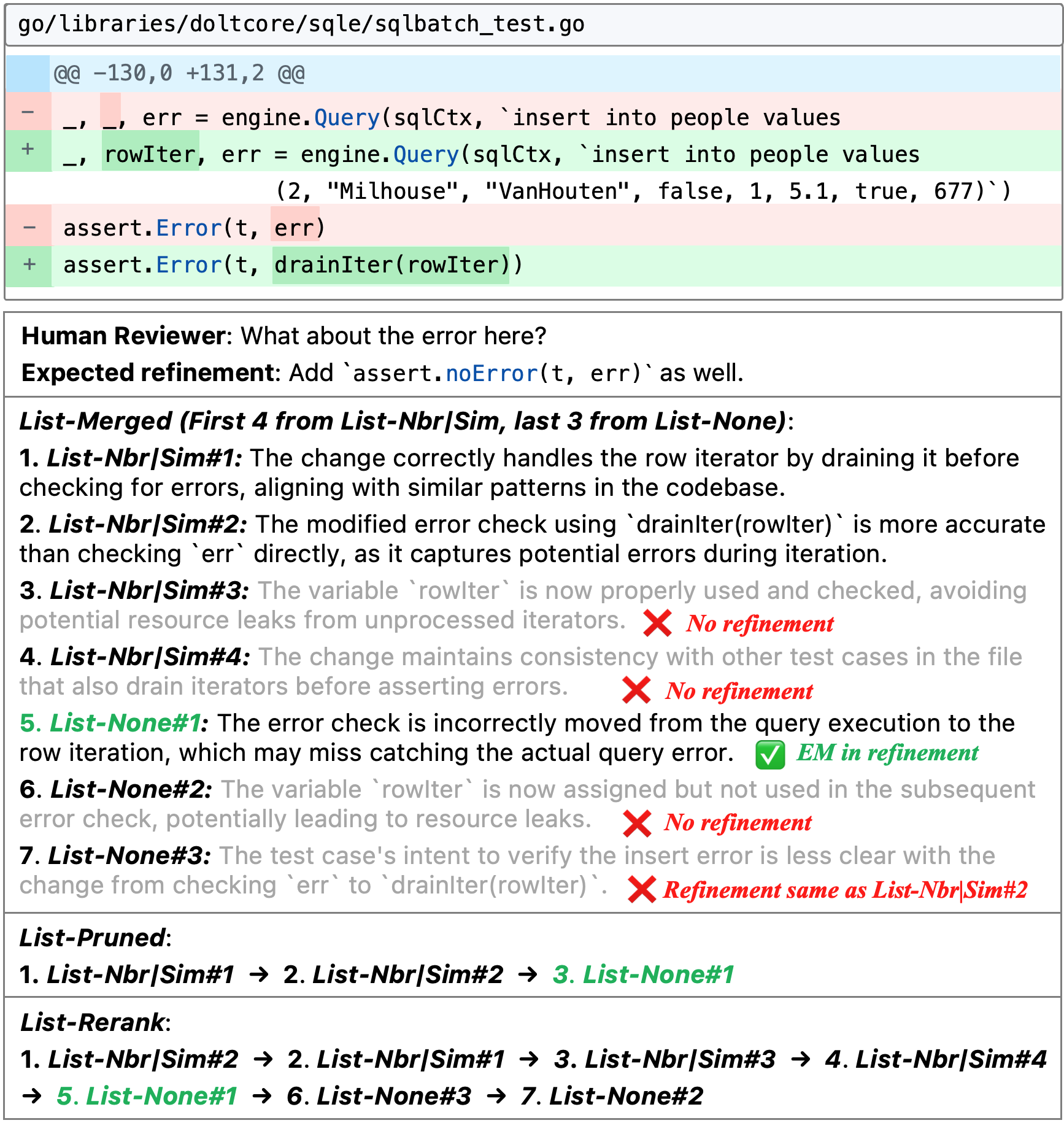}
	\caption{An example illustrating how different candidate integration strategies reorganize the merged candidate list.}
	\label{fig:rq3_example}
\end{figure}

% This indicates that candidates generated with and without contextual augmentation remain complementary.
% This complementarity is consistent with our RQ2 findings.
The complementarity between context-enhanced and no-context generation is consistent with our RQ2 findings.
Additional context can redirect the model's attention, so even the best context-enhanced setting may still miss basic issues that are recoverable without extra context.
Figure~\ref{fig:rq3_example} shows such a \href{https://github.com/dolthub/dolt/pull/869#discussion_r487469389}{case} from \texttt{dolthub/dolt}.
The human reviewer points out that the revised code should still assert on \texttt{err}.
This issue is correctly surfaced and ranked first by \sit{List-None}.
In contrast, the four comments from \sit{List-Nbr\textbar Sim} only justify the modified code and do not provide any effective review suggestion.
This example suggests that retaining the no-context candidates as a fallback remains useful, as context may sometimes rationalize the current change and suppress issues that would otherwise be detectable.

\begin{table*}[t]
\caption{\textbf{Ablation Study.} Each row removes one component from the full pipeline (\sit{List-Pruned}). Each variant is evaluated on the N cases for which at least one candidate is retained. $\triangle$ values are relative to the full approach.}

\label{tab:ablation}
\centering
\scriptsize
\begin{tabular}{cccccccc}
\hline
\textbf{Variant} & \textbf{N} & \textbf{Avg. Cands} & \textbf{EM Cases} & \textbf{RefineEM} & \textbf{RefineBLEU} & \textbf{ReviewBLEU} & \textbf{ReviewBERT} \\
\hline
\textbf{Full} & 1,408 & 3.4 & 402.7 & 28.60\% & 84.72 & 40.40 & 5.69 \\
\hline
\textbf{w/o taxonomy}       & 1,398 & 2.9 (-0.5) & 388.7 (-14.0) & 27.80\% (-0.80) & 84.35 (-0.37) & 39.79 (-0.61) & 5.63 (-0.06) \\
\textbf{w/o workflow}       & 1,398 & 3.2 (-0.2) & 398.3 (-4.3)  & 28.50\% (-0.10) & 84.54 (-0.18) & 40.22 (-0.18) & 5.78 (+0.09) \\
\textbf{w/o issue-list}     & 1,310 & 1.8 (-1.6) & 310.0 (-92.7) & 23.67\% (-4.93) & 81.45 (-3.27) & 39.00 (-1.40) & 5.74 (+0.05) \\
\textbf{w/o context augmentation} & 1,380 & 2.8 (-0.5) & 353.0 (-49.7) & 25.58\% (-3.02) & 83.56 (-1.16) & 39.43 (-0.97) & 5.58 (-0.11) \\
\textbf{w/o merging}        & 1,372 & 2.4 (-1.0) & 368.0 (-34.7) & 26.83\% (-1.77) & 83.43 (-1.29) & 38.65 (-1.75) & 5.39 (-0.30) \\
\textbf{w/o pruning}        & 1,438 & 7.2 (+3.9) & 402.7 (+0.0)  & 28.00\% (-0.60) & 85.65 (+0.93) & 44.36 (+3.96) & 6.31 (+0.62) \\
\hline
\end{tabular}
\end{table*}

However, direct merging inevitably enlarges the candidate pool, with \sit{List-Merged} producing 7.2 comments per test case on average, which necessitates a further organization step.
Of the two strategies we compare, our refinement-guided pruning strategy (\sit{List-Pruned}) is clearly more effective.
As shown in Table~\ref{tab:rq3-results}, it reduces the average candidate count from 7.2 to 3.4 while improving \textit{EM Cases} and \sit{RefineEM} at every cutoff, although slightly lowering the review-text metrics.
In particular, \sit{List-Pruned}@5 preserves a practical cutoff: with an average of 3.1 candidates, it achieves 401.0 EM Cases — 99.6\% of the full-list ceiling — while avoiding the excessively long cases, which can still reach 11 comments after pruning.
Moreover, pruning removes all candidates for 30 cases (out of 1,438).
Rather than treating this as a failure, we interpret it as the system choosing not to produce low-value comments, which is often a desirable behavior in practice.
Figure~\ref{fig:rq3_example} further illustrates this effect.
Pruning removes \sit{List-Nbr\textbar Sim}\#3, \sit{List-Nbr\textbar Sim}\#4, and \sit{List-None}\#2 because they do not induce any code revisions, and removes \sit{List-None}\#3 because it leads to the same refinement as \sit{List-Nbr\textbar Sim}\#2.
As a result, the candidate list shrinks from seven comments to three, and the human-aligned comment moves from rank~5 to rank~3.
By contrast, although \sit{List-Rerank} slightly improves some review-text metrics at the same cutoff, it underperforms \sit{List-Merged} in both \textit{EM Cases} and \sit{RefineEM}.
In the case of Figure~\ref{fig:rq3_example}, the reranking method still places the non-actionable \sit{List-Nbr\textbar Sim} comments ahead of the human-aligned \sit{List-None}\#1 comment, despite explicitly considering dimensions such as actionability.

These results suggest that the quality dimensions used by Lu et al.\ may not fully capture what matters for this task.
Those dimensions largely evaluate how well a comment presents itself along axes such as readability and specificity.
However, as illustrated in Figure~\ref{fig:rq1_example}, LLMs already excel at producing readable and specific comments, often surpassing the supervised baseline and, in some cases, even being more concrete than human reviewers.
What is needed instead is issue priority rather than comment quality, i.e., among multiple plausible comments, which one targets the most important issue and is most likely to induce a meaningful code revision.
A promising direction is therefore to incorporate human review cases as references for priority.
Action-based evaluation through simulated refinement is also better aligned with this goal than direct LLM-based comment scoring.
Our refinement-guided pruning strategy is an initial attempt in this direction.

\begin{rqanswerbox}
\textbf{Answer to RQ3:}
Integrating candidates from context-enhanced and no-context review (\sit{List-Merged}) yields \textbf{34.7} more EM cases and improves \sit{RefineEM} by \textbf{2.41} pp over the best single setting, reaching 402.7 EM Cases (28.00\% \sit{RefineEM}), at the cost of a larger candidate pool (7.2 per case on average).
Refinement-guided pruning (\sit{List-Pruned}) effectively reduces this to \textbf{3.1} candidates at top-5 while retaining nearly the full refinement benefit.
\end{rqanswerbox}

%%%%%%%%%%%%%%%%%%%%%%%%%%%%%%%%%%%%%%%% Ablation Study %%%%%%%%%%%%%%%%%%%%%%%%%%%%%%%%%%%%%%%%
\subsection{Ablation Study}
\label{sec:ablation}

To clarify the contribution of each design choice, we conduct an ablation study on our full pipeline (\sit{List-Pruned}), which performs issue-list generation guided by a defect taxonomy and a structured review workflow, merges candidates from the best context-enhanced (\sit{List-Nbr\textbar Sim}) and no-context setting (\sit{List-None}), and applies refinement-based pruning.
We remove each component individually and report results in Table~\ref{tab:ablation}.
For \textit{w/o taxonomy}, we remove the defect taxonomy from the prompt while keeping the review workflow and issue-list format.
For \textit{w/o workflow}, we remove the review workflow guidance while keeping the taxonomy and issue-list format.
For \textit{w/o issue-list}, we replace issue-list generation with primary-issue generation under the same two context settings (\sit{Primary-Nbr\textbar Sim} and \sit{Primary-None}), then merge and prune.
For \textit{w/o context augmentation}, we replace \sit{List-Nbr\textbar Sim} with a second \sit{List-None} run, so both merged candidates come from no-context generation.
For \textit{w/o merging}, we use only \sit{List-Nbr\textbar Sim} candidates without merging in the no-context run, and apply pruning directly.
For \textit{w/o pruning}, we remove the refinement-based pruning step and retain the full merged candidate pool.

As shown in Table~\ref{tab:ablation}, removing issue-list generation causes the largest drop ($-4.93$ pp \sit{RefineEM}, $-92.7$ EM cases), indicating that generating multiple candidate issues is a necessary condition for other designs such as context enhancement and candidate merging to take effect.
Context augmentation contributes more than merging alone: removing additional context causes larger declines than removing merging in \sit{RefineEM} ($3.02$ pp vs.\ $1.77$ pp) and EM cases (49.7 vs.\ 34.7), indicating that stronger candidate quality from context enhancement matters more than simply combining candidates from multiple runs.
Among the prompt components, the defect taxonomy contributes more than the review workflow to surfacing a broader range of review concerns ($-0.80$ pp vs.\ $-0.10$ pp in \sit{RefineEM} when removed).
Finally, as previously discussed, refinement-based pruning mainly contributes by controlling the candidate set size.
Overall, the design choices contribute in complementary ways to the full pipeline.

\subsection{Implications}

Based on our findings, we discuss several implications for practitioners who want to deploy the LLM-based code review and researchers studying LLM-based code review.

\vspace{1ex}
\textbf{For Practitioners:}

\textit{1) Enumerate issues, but do not trust the model's ranking.}
Our results suggest that current LLMs can identify a broad range of plausible review issues, but their prioritization does not reliably align with human reviewers.
Thus, the developers could leverage the model for broad issue discovery, but should remain critical about the priorities it assigns.

\textit{2) Prefer selective context to more context.}
Contextual augmentation is useful, but more context is not always better: blindly adding context can distract the model from issues it would otherwise catch while also increasing token cost. 
Among the strategies we evaluated, combining neighboring context with similar co-change context yields the strongest gains and appears to be a practical augmentation choice. 
Its benefit also depends on prompt design: richer context is better supported by a detailed prompt with explicit defect taxonomies and review guidelines, while lighter context settings can remain competitive under simpler prompts.
Meanwhile, our candidate-merging results suggest that a lightweight no-context pass is worth retaining by default, since it can recover issues that context-enhanced generation may suppress.

\textit{3) Refinement feedback can improve both comment filtering and presentation.}
Our results show that refinement-guided pruning is an effective way to control candidate explosion in multi-candidate generation.
Beyond filtering, the predicted code revision could also be surfaced alongside the comment itself. This would turn interaction with the reviewer into a selection task: the developer sees not only a natural-language suggestion, but also the concrete code change it implies, making candidate evaluation easier and the review process more actionable.

\vspace{1ex}
\textbf{For Researchers:}

\textit{1) Issue prioritization as an open problem.}
Our findings point to issue prioritization as an important open problem in LLM-based code review.
The main limitation is no longer simply whether the model can notice a potential issue, but whether it can judge which issue matters most in the review context.
Future work should pay more attention to priority modeling, e.g. leveraging historical review cases as supervision for deciding which concerns deserve to be surfaced first.

\textit{2) Context organization beyond retrieval.}
Our results suggest that retrieving more context does not guarantee better reviews.
This suggests that context selection, pruning, and presentation are also important research challenges alongside retrieval itself.
A promising direction is to move beyond flat context concatenation.
Different context configurations could be assigned to independent sub-agents, each generating suggestions with its own supporting evidence.
Their outputs could then be compared or jointly deliberated, allowing conflicting signals to be weighed more explicitly than when all context is merged into a single prompt.

\textit{3) Action-grounded evaluation as a feedback loop.}
Text-similarity metrics such as BLEU are insufficient for assessing the real quality of review comments, because they measure surface form rather than practical impact.
Beyond measurement, grounded evaluation is also important because it provides a concrete feedback signal for improving LLM-based code review.
When a model receives an action-based response, such as whether its suggested revision can actually be carried out and whether the resulting code is correct, it can adapt its behavior accordingly.
Such consequence-level feedback is difficult to obtain from text-similarity scores alone.

%%%%%%%%%%%%%%%%%%%%%%%%%%%%%%%%%%%%%%%% Threats to Validity %%%%%%%%%%%%%%%%%%%%%%%%%%%%%%%%%%%%%%%%
\section{Threats to Validity}

\vspace{1ex}
\textbf{Internal Validity.}
One internal threat in our study is from using refinement-based evaluation as the primary effectiveness metric.
We use CodeReviewer~\cite{li2022automating} to simulate the code changes that a generated review comment may induce, and compare the predicted revision against the final human revision.
This introduces potential bias, because the refinement model may occasionally produce a correct revision for the wrong reason, or fail to realize a valid review suggestion even when the underlying issue is correctly identified.
To mitigate this, we apply the same CodeReviewer-based refinement pipeline consistently across all evaluated settings, so the comparison remains fair even if the simulator itself is imperfect.
Furthermore, Pornprasit et al.~\cite{pornprasit2024fine} show that CodeReviewer's refinement performance on this dataset is second only to fine-tuned GPT-3.5, outperforming GPT-3.5 without fine-tuning, making it the strongest available off-the-shelf refinement model for this task.
A second internal threat is the non-determinism of LLM generation.
Although we use temperature~0, model outputs may still vary across runs.
To reduce this effect, we run each setting three times and report the averaged results.

\vspace{1ex}
\textbf{External Validity.}
Our evaluation is limited to Go review instances.
The effectiveness of contextual augmentation may differ in other programming languages, especially those with different language characteristics or tooling ecosystems, such as Python with more dynamic typing or Java with heavier build and dependency structures.
A second external threat is that we evaluate only a single LLM, DeepSeek-V3.
Different models may vary in their ability to follow review instructions, process long context, and prioritize issues, which could affect the relative benefits of issue-list review and contextual augmentation.
Finally, our dataset covers 54 open-source Go repositories from GitHub and may not represent industrial codebases or review practices in other ecosystems.
Replicating our approach across additional LLMs, programming languages, and review environments is an important direction for future work.
For similar-context retrieval, we restrict the search to co-modified files to obtain a task-coupled and high-precision search space, rather than a broad repository-wide pool.
Extending retrieval to the full repository snapshot would introduce a substantially noisier candidate pool and require indexing the entire codebase at each commit; we leave this for future investigation.

%%%%%%%%%%%%%%%%%%%%%%%%%%%%%%%%%%%%%%%% Related Work %%%%%%%%%%%%%%%%%%%%%%%%%%%%%%%%%%%%%%%%
\section{Related Work}
In this section, we discuss prior research on automated AI-based code review and its evaluation.

\vspace{1ex}
\textbf{Automated AI-based Code Review.}
Automating code review has attracted growing attention with the rise of generative AI.
Prior work has organized this task into three core tasks~\cite{lin2025codereviewqa}:
(1) determining whether a code change needs review~\cite{hellendoorn2021towards};
(2) generating natural-language review comments~\cite{li2022automating, tufano2022using, li2022auger, Lin2024};
and (3) generating revised code based on the comments~\cite{li2022automating, pornprasit2024fine, abtahi2025augmenting}.
Among these, generating comments is particularly challenging, as it requires reasoning about code semantics and producing context-sensitive suggestions.
Leveraging the large volumes of review data accumulated in open-source communities, early approaches trained encoder-decoder models such as T5~\cite{tufano2022using} to generate review comments, while later work explored task-specific pretraining~\cite{li2022auger,li2022automating} or fine-tuning of general-purpose LLMs like LLaMA~\cite{lu2023llama, yu2024fine}.

Recent studies have further investigated how to apply LLMs to code review more effectively.
Zhang et al.~\cite{zhang2025laura} leverage pull request metadata, historical review comments, and change-related functions (akin to our neighborhood context) to support more reliable review generation.
Peng et al.~\cite{peng2025icodereviewer} propose a mixture-of-prompts architecture that dynamically routes the input program to activate only the most relevant prompt experts, achieving diverse yet precise security-focused reviews.
In industry, companies such as ByteDance~\cite{sun2025bitsai} combine traditional static analysis, heuristic rule networks, and LLMs to enhance the reliability of LLM-generated code reviews.

Compared with these studies, our work focuses on two questions that remain underexplored:
how different forms of code context, especially non-local context obtained through LSP-based navigation and IR-based retrieval, affect LLM-based code review; and how different generation paradigms (e.g., issue-list review merged with refinement-guided filtering) affect LLM-based code review performance.

\vspace{1ex}
\textbf{Effectiveness Metrics for Code Review Generation.}
Prior work~\cite{li2022automating, tufano2022using, naik2025crscore} largely relied on textual similarity metrics, such as BLEU or embedding-based scores, to measure how closely model-generated comments match human references.
However, such superficial metrics fail to capture the practical value of review comments.

A growing body of recent work has therefore sought more realistic evaluation frameworks.
Lu et al.~\cite{lu2025deepcrceval} propose an LLM-as-a-judge framework that assesses generated comments along multiple quality dimensions, including readability and actionability.
Pereira et al.~\cite{pereira2026cr} compare AI-generated comments against the final bug-fixing patch to determine whether the comment correctly identified the real issue.
Zhang et al.~\cite{zhang2026code} introduce a test-based benchmark to measure the real-world utility of AI-generated reviews.

Our study shares this broader goal of going beyond superficial text similarity to evaluate the practical value of generated review comments.
Instead of comparing comments only by their similarity to human-written feedback, we evaluate them by the code revisions they induce against the actual human revision.
This provides a more action-grounded measure of review effectiveness.

%%%%%%%%%%%%%%%%%%%%%%%%%%%%%%%%%%% Conclusion and Future Work %%%%%%%%%%%%%%%%%%%%%%%%%%%%%%%%%%%
\section{Conclusion and Future Work}
In this paper, we studied improving LLM-based code review through issue-list generation, contextual augmentation, and refinement-guided candidate integration.
On 1,438 Go review instances, we showed that 
issue-list generation is more effective than primary-issue review, 
combining neighboring context with similar co-change context is the strongest context setting we evaluated, 
merging context-enhanced and no-context candidates further improves coverage
and refinement-guided pruning can substantially reduce candidate-set size while retaining nearly the full refinement benefit.
We further summarize our findings as several implications for practitioners and researchers, which also suggest future directions including issue prioritization, context organization beyond retrieval, and action-grounded evaluation as a feedback loop for improving LLM-based code review.

%%%%%%%%%%%%%%%%%%%%%%%%%%%%%%%%%%%%%%%% Appendix %%%%%%%%%%%%%%%%%%%%%%%%%%%%%%%%%%%%%%%%
% if have a single appendix:
%\appendix[Proof of the Zonklar Equations]
% or
%\appendix  % for no appendix heading
% do not use \section anymore after \appendix, only \section*
% is possibly needed

% use appendices with more than one appendix
% then use \section to start each appendix
% you must declare a \section before using any
% \subsection or using \label (\appendices by itself
% starts a section numbered zero.)
%

% \appendices
% \section{Proof of the First Zonklar Equation}
% Appendix one text goes here.

% % you can choose not to have a title for an appendix
% % if you want by leaving the argument blank
% \section{}
% Appendix two text goes here.

%%%%%%%%%%%%%%%%%%%%%%%%%%%%%%%%%%%%%%%% Acknowledgment %%%%%%%%%%%%%%%%%%%%%%%%%%%%%%%%%%%%%%%%
% use section* for acknowledgment
% \section*{Acknowledgment}
% The authors would like to thank...

%%%%%%%%%%%%%%%%%%%%%%%%%%%%%%%%%%%%%%%% Reference %%%%%%%%%%%%%%%%%%%%%%%%%%%%%%%%%%%%%%%%
% Can use something like this to put references on a page
% by themselves when using endfloat and the captionsoff option.
\ifCLASSOPTIONcaptionsoff
  \newpage
\fi

% trigger a \newpage just before the given reference
% number - used to balance the columns on the last page
% adjust value as needed - may need to be readjusted if
% the document is modified later
%\IEEEtriggeratref{8}
% The "triggered" command can be changed if desired:
%\IEEEtriggercmd{\enlargethispage{-5in}}

% references section
\bibliographystyle{IEEEtran}
\bibliography{refs}

\end{document}